\documentclass[11pt]{article}

\usepackage[nocompress]{cite}
\usepackage[utf8]{inputenc}
\usepackage{amsmath}
\usepackage{hyperref}
\usepackage{amsfonts}
\usepackage{graphicx}
\usepackage{bm}
\usepackage{pifont}
\usepackage[linewidth=1pt]{mdframed}
\usepackage{caption}
\usepackage{subcaption}
\usepackage{algorithm}
\usepackage[noend]{algpseudocode}
\usepackage{array}

\advance \oddsidemargin by -0.8in
\newcommand{\myparskip}{3pt}
\parskip \myparskip

\begin{document}
\title{User-Guided Free-Form Asset Modelling}
\author{Daniel Beale \thanks{University of Bath. Email: {\tt d.beale@bath.ac.uk}. Supported by the EPSRC grant EP/K02339X/1 - "Acquiring Complete and Editable Outdoor Models from Video and Images" }}

\begin{titlepage}
\maketitle

\thispagestyle{empty}

\begin{abstract}
In this paper a new system for piecewise primitive surface recovery on point clouds is presented, which allows a novice user to sketch areas of interest in order to guide the fitting process. The algorithm is demonstrated against a benchmark technique for autonomous surface fitting, and, contrasted against existing literature in user guided surface recovery, with empirical evidence. It is concluded that the system is an improvement to the current documented literature for its visual quality when modelling objects which are composed of piecewise primitive shapes, and, in its ability to fill large holes on occluded surfaces using free-form input.
\end{abstract}

\end{titlepage}

\section{Introduction}
Creating models of existing buildings and man made structures is a problem for a range of different industries such as visual effects, gaming and architecture, and is one of the most important challenges for computer graphics. It is possible for highly skilled artists to reconstruct an existing building if reference photographs and plans are available. This process can be expensive, motivating the search for alternatives that build models directly from image or video data with little or no demand placed on the user.

The need for 3D models from the real world has lead to a wide variety of research and systems. Previous work in this field is not far removed from the paradigm of traditional geometric modelling packages such as Maya\footnote{\url{http://www.autodesk.co.uk/products/maya/overview}} and Modo\footnote{\url{https://www.thefoundry.co.uk/products/modo/}}; the user, at a conventional workstation, is able to perform standard operations such as selecting specific points and drawing precise curves. While this workflow leads to models which are visually pleasing from the point of view of an artist or director, they do not easily model the geometry of a real existing scene, leaving room for techniques for speeding up the process through automation.

Standard industrial reconstruction pipelines recreate surfaces from photographs by first building a point cloud, which can be obtained from a number of different technologies such as photographs or range scanners. A common problem among all of these technologies is the un\-desired production of holes in regions which are occluded, for example, or exhibit a high degree of specular reflection or, indeed, no reflection at all. Autonomous methods for filling holes rely on prediction and prior knowledge of the expected surface shape. User guided techniques are able to draw on human prior experience for identifying surfaces, but also require a way to interact with the system in a way which is not time consuming or difficult.

Rather than motivating the problem through the need for precise models, attention is drawn to the necessity of accurate and fast modelling  using current point cloud technology. The demands of modern film production, and increasingly television and gaming, requires pre-visualisation\footnote{\url{https://www.smpte.org/sections/hollywood/previsualization-current-filmmaking}}\footnote{\url{http://digitalfilmfarmworkshops.com/pre-visualize-directing-video-film/}} ; this is the task of rapidly prototyping and visualising the individual shots of a film before actual production. 
While pre-visualization has clear demands from a technical point of view; including the need for logistical planning, safety, rigging and setup, and managing costs; there are also many artistic advantages, for example, ensuring that you get the shots you want during production. With the rise of virtual production, the turn-around times between pre-visualisation and production are getting smaller. In the area of 3D modelling, these time pressures give rise to a need for rapid prototyping of geometry that can be used to visualise and plan shots before subsequent detailed modelling for final post-production and composting.

The need for methods which allow a user to quickly build surface models from point clouds are identified, while maintaining robustness to large holes and noise. The new system introduced in this paper allows a user to guide fitting through a collection of free form curves. The workflow first requires the user to highlight required points on the cloud, and then fit a collection of maximum-likelihood primitive surfaces over a number of user interactions. The final surfaces are closed using a well-known and contemporary iso-surface algorithm. It is concluded that the method presented in this paper produces high quality surfaces, and is an improvement beyond contemporary methods when compared on a dataset of man-made objects. 

\section{Background}
Reconstructing point clouds and surfaces has motivated the fields of computer vision and graphics for many years, and remains an area of active interest. Methods are developed which remain accurate in the presence of noise and occlusion, maintain a low algorithmic complexity and also allow creative but precise user input. 

Since surface reconstruction has been successfully used in industry for some time now, the common types of software pipeline are detailed with allusion to creative contributions in the area. A large number of algorithms are documented in order to solve specific problems that a general reconstruction algorithm cannot. To begin with, major, fully autonomous systems for surface recovery from images are detailed, and also, alternative hardware such as range scanners. Methods for improving the process through user guidance are introduced towards the end of the section. 

There is a large body of research in to surface reconstruction from point clouds, with algorithms that are used in modelling software such as The Foundry's Nuke\footnote{\url{https://www.thefoundry.co.uk/products/nuke/}}. The area has a well formed and standard pipeline, which is useful for the range of different surfaces that it can deal with, and also for its ability to process point clouds which are generated from alternative hardware. The pipeline consists of the following processes,
\begin{enumerate}
	\item \textbf{Point cloud reconstruction}. A dense collection of points is reconstructed from images using available software e.g.\cite{vsfm1,vsfmba,pmvs,kolmogorov2004graph}. Alternatively one can use a different sensor such as a range finder, radar or sonar with varying reconstruction qualities.
	\item \textbf{Normal estimation}. A consistent set of normals are estimated for each point in the cloud. This indicates both where the surface is and also which side of the surface a point is on \cite{mitra2003estimating,hoppe1992surface}.
	\item \textbf{Surface fitting}. Finally, an implicit surface; such as a Poisson surface \cite{kazhdan2006poisson,kazhdan2013screened,bolitho2009parallel}, a Gaussian Process \cite{williams2007gaussian,dragiev2011gaussian} or Delaunay triangulation \cite{gopi2000surface}; or a parametric surface such as a B-Spline \cite{krishnamurthy1996fitting,wang2006fitting} is fit to the data.
\end{enumerate}

A main alternative to the method described above is to reconstruct a volume or surface directly from a collection of images, without the intermediate point cloud reconstruction. These methods produce excellent results when there are a large number of cameras viewing the object (e.g. ~\cite{furukawa2009reconstructing,campbell2010automatic,starck2006volumetric}), but will not work when there are only a few images. They are a good candidate for reconstructing models of objects which are small enough to be viewed from a large number of angles, but are generally unsuitable for medium to large scale objects such as buildings and landscapes.

The primary concern of this paper is with man made objects, on which the assumption of `piecewise primitive' surface structure is reasonable.  It is beneficial to assume that the surface is piecewise primitive if there are large holes in the point cloud or it is extremely noisy, since the assumptions places a large enough constraint that it is possible to predict with a high degree of accuracy in unknown areas. There are methods which are able to fit primitives such as toroids, quadrics, or super-quadrics without user interaction~\cite{debevec1996modeling,musialski2013survey,bo2012revisit}. For example, \cite{monszpart2015rapter} fit a collection of planes to the surface using an energy minimisation framework; \cite{schnabel2007efficient} use probabilistic RanSaC to fit primitives including toroids and cylinders; \cite{li2011globfit} optimise using a formulation which minimises global and pairwise energy; while \cite{labatut2009hierarchical} minimise energy using a binary space partitioning tree. For man made objects this approach is reasonable, and well studied, but complicated natural phenomena, such as tree bark or moss, break the assumptions and so produce poor quality results.

Incorporating the use of a small amount of human interaction can be a large benefit to the process (e.g. \cite{Hengel:2007aa,portelli2010framework}). One can provide extra points and normals to surface by selecting a plane which is orthogonal to a point on the medial axis and sketching on it~\cite{yin2014morfit}, before sweeping along the skeleton to fit the surface. Methods which fit a collection of planes to the surface are useful since they easily snap together along the edges \cite{sinha2008interactive,nan2010smartboxes}. Using quadric fitting with interaction has been studied in~\cite{andrews2014type}. The authors allow a user to identify a collection of points on a mesh from a single stroke and fit a quadric, displaying triangles which are near the original mesh. \cite{cao2014interactive} consider objects which are made by extruding 2D profiles, such as furniture and buildings. The work described in~\cite{habbecke2012linear} use software to apply constraints to designs that are made against pictures, but do not use computer vision in the fitting process.

Alternative to the multiple image / 3D-reconstruction pipeline, shape and volume can be quickly recovered from a single image, without any knowledge of the scene geometry ~\cite{lau2010modeling,Chen:2013aa,jiang2009symmetric}. In 3-Sweep~\cite{Chen:2013aa} a user can quickly outline contours on objects and recover depth using high level knowledge about object ordering, the resulting models can then be recomposed into different photographs. Methods in this area can quickly recover plausible looking geometry. Without the addition of strong, and potentially unavailable, priors, they do not have the same metric accuracy as a point cloud reconstruction. 

It is worth mentioning that editable assets can also be generated from scratch using interaction alone, without any computer vision (e.g.~\cite{chen2008sketching,lipp2014pushpull++,xu2014true2form,iarussi2015bendfields}).
These family of approaches are suited to modelling novel buildings from an architectural standpoint, and belong to a large class of CAD modelling software. The prior use of point-based reconstruction provides a cue to speed up the extraction of surfaces on existing buildings. It should be viewed as a potential aid for current CAD modelling tools, removing the need for intensive user interaction where a 3D model already exists.

\begin{table}[h!]
	\centering
	\def\arraystretch{1.5}
	\begin{tabular}{ | r | c | c |} \hline
		& \textbf{Fit type} & \textbf{Input}
		\\ \hline
		\textbf{Sinha et al.} & Planes & Polygon tool
		\\ \hline
		\textbf{VideoTrace} & B-Splines & Spline tool
		\\ \hline
		\textbf{Morphit} & B-Splines & Free-form
		\\ \hline
		\textbf{This system} & Quadric forms & Free-form \\ \hline
	\end{tabular}
	\caption{A comparison of some relevant, and current, interactive techniques for surface reconstruction.}
	\label{tab:litcompare}
\end{table}
\begin{figure*}[t]
\centering
\includegraphics[width=0.8\textwidth]{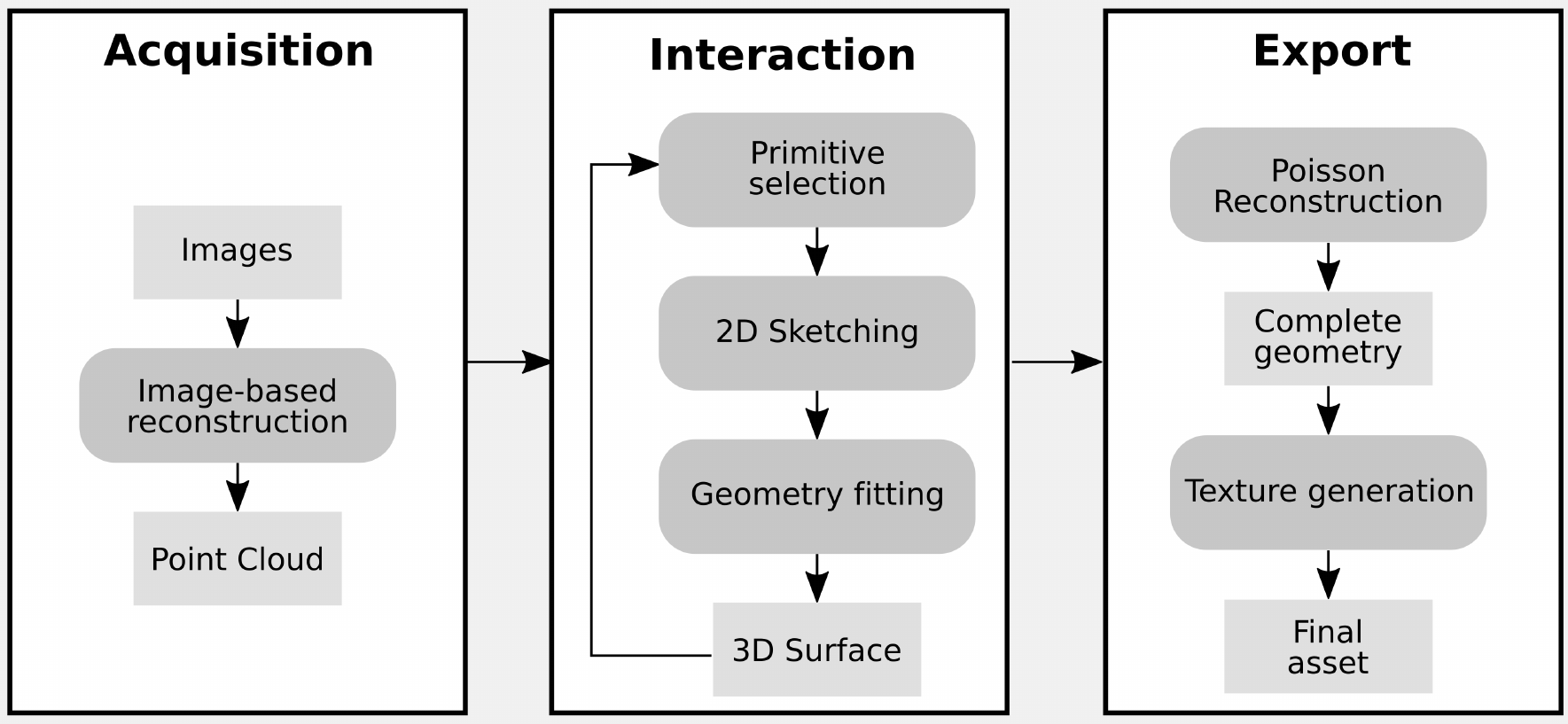}
\caption{A schematic diagram of the system.}
\label{fig:scheme}
\end{figure*}

The system presented in this paper follows the standard modelling pipeline described at the beginning of this section. A point cloud is reconstructed first, before any user interaction, to extract points for fitting geometric primitives. This places the work in to a mainstream research area, allowing it to be compared to fully autonomous point cloud techniques, and also interactive multiple view surface reconstruction algorithms. 

The literature arranged in Table \ref{tab:litcompare} is a representative sample of the current state of the art in interactive surface recovery. It compares the system presented in this paper to the most recent and most similar alternative systems in the area. It can be seen that while \cite{sinha2008interactive} and VideoTrace ~\cite{Hengel:2007aa} are able to fit primitive parametric models to the data, they cannot deal with free-form input. The method in Morphit ~\cite{yin2014morfit} allows a user to improve implicit surface fitting using free-form input, which works well on complicated surfaces but cannot predict when there are large holes in the surface. While there does not appear to be any previous work that uses rough, or loose, annotations for primitive geometry recovery, there are other applications that have similar interfaces for related editing tasks such as colorization~\cite{Levin2004} and segmentation~\cite{Subr2013}.

The limitations of contemporary work is given a more detailed review in the results section, with some specific examples and a more in depth discussion. It is concluded that the method presented in this paper is a contribution to the area of free-form modelling with geometric primitives.

\section{System Overview}
\begin{figure*}[t!]
	\centering
	\def\arraystretch{2}
	\begin{tabular}{  >{\centering\arraybackslash}m{0.1in} >{\centering\arraybackslash}m{3.5in} >{\centering\arraybackslash}m{0.1in} >{\centering\arraybackslash}m{1.4in}}
	 & \textbf{Point selection} & & \textbf{Surface fitting} \\
	\rotatebox{90}{\hspace{0in}\textbf{Quadric} }& 
	\includegraphics[width=0.25\textwidth]{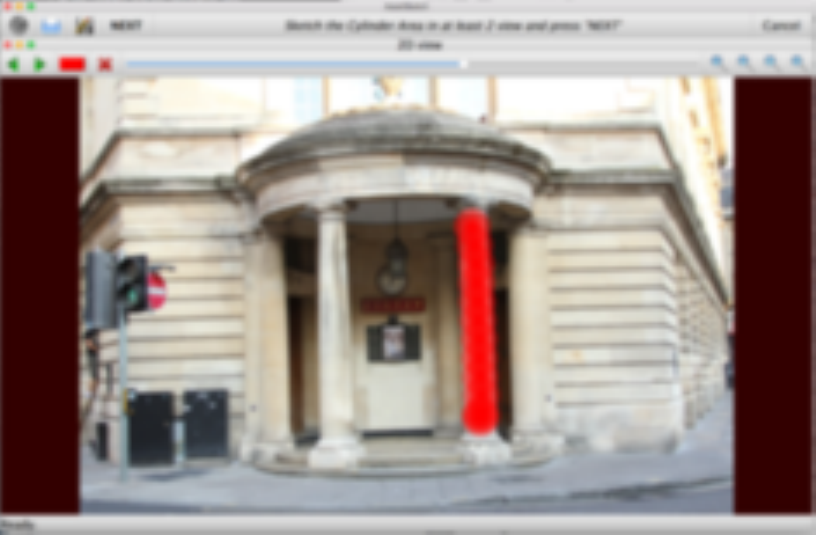} 
	\includegraphics[width=0.25\textwidth]{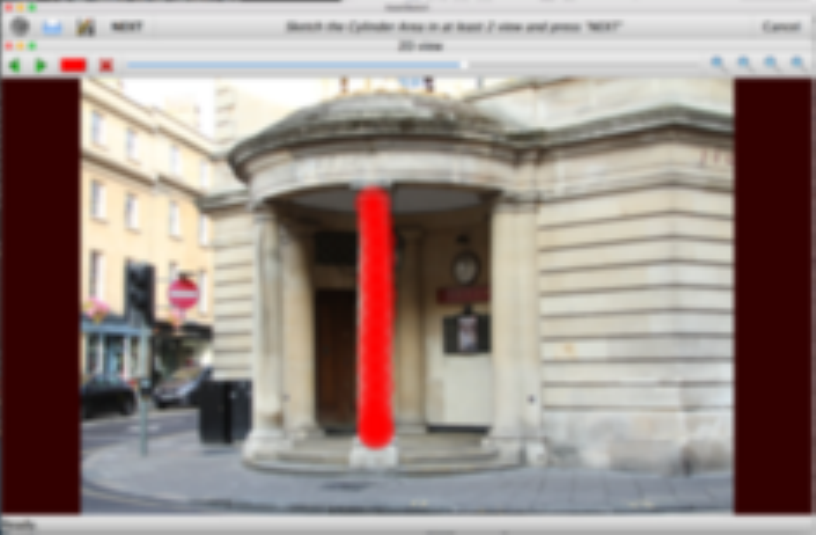} & $\rightarrow$  & 
	\includegraphics[width=0.25\textwidth]{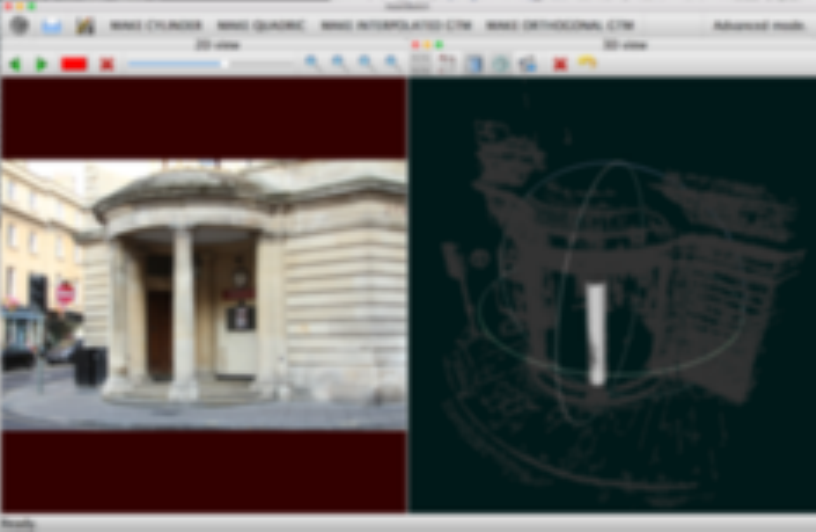} \\
	\rotatebox{90}{\hspace{0in} \textbf{LVM} }& 
	\includegraphics[width=0.25\textwidth]{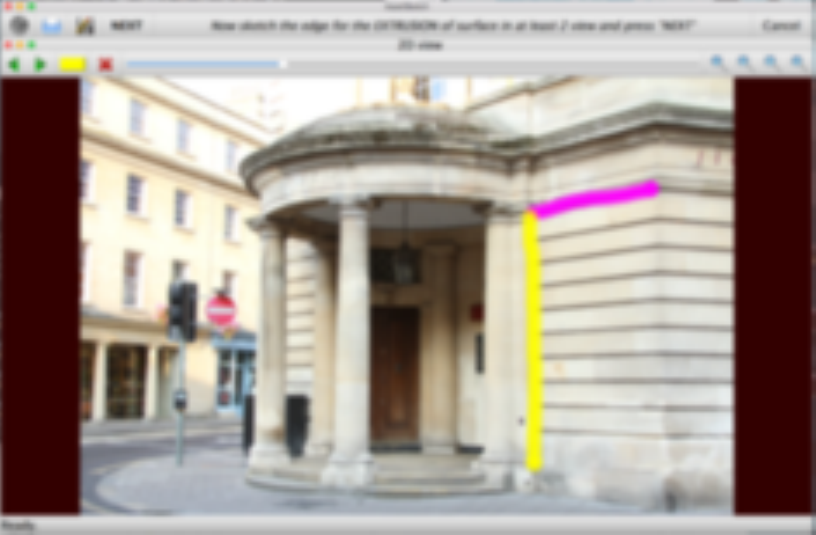} 
	\includegraphics[width=0.25\textwidth]{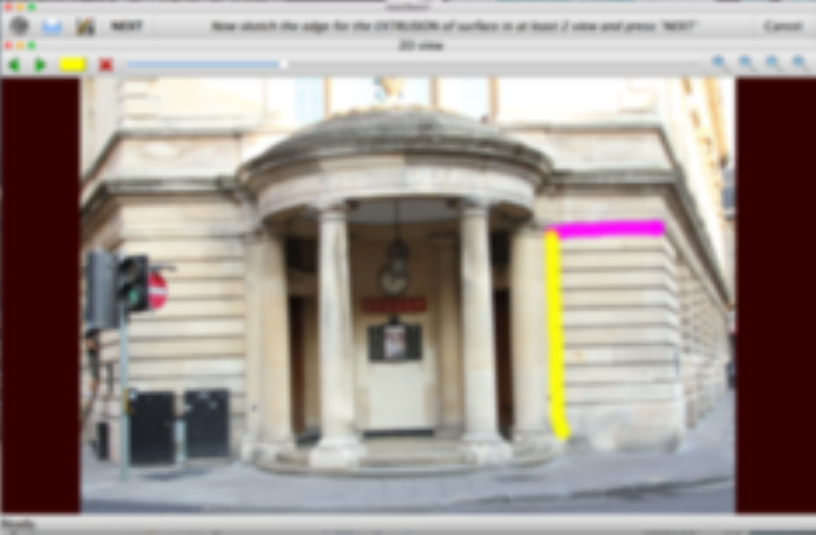} & $\rightarrow$  & 
	\includegraphics[width=0.25\textwidth]{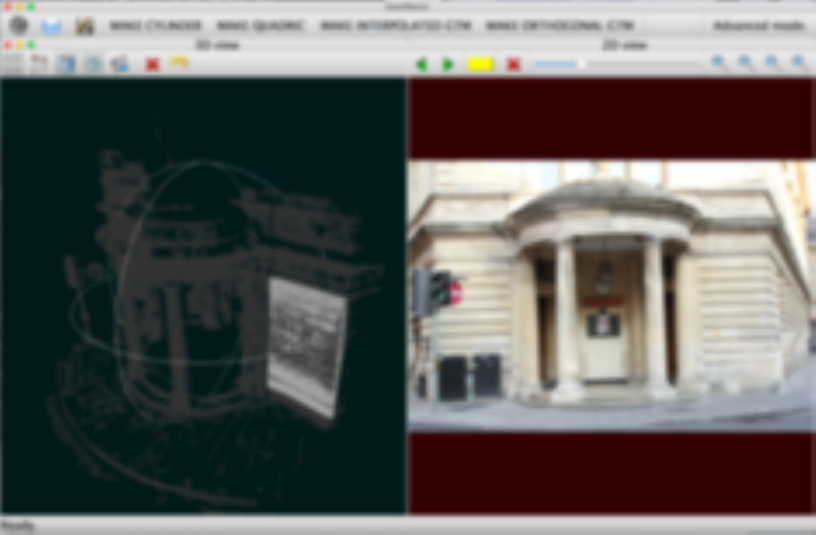} 
	\end{tabular}
	\caption{An example of some sketch input to the system and the corresponding output from the primitive fitting algorithms. The first row contains an example of fittinAn example of someg an ellipsoid to a column of a building, and the second row is an example of a perpendicular extrusion using a pair of latent variable models.}
	\label{fig:examplesketchinput}
\end{figure*}
\begin{figure}[t!]
	\centering
	\def\arraystretch{3}
	\begin{tabular}{  c c }
	Primitive geometry & Poisson reconstruction \\
	 \includegraphics[width=0.35\textwidth]{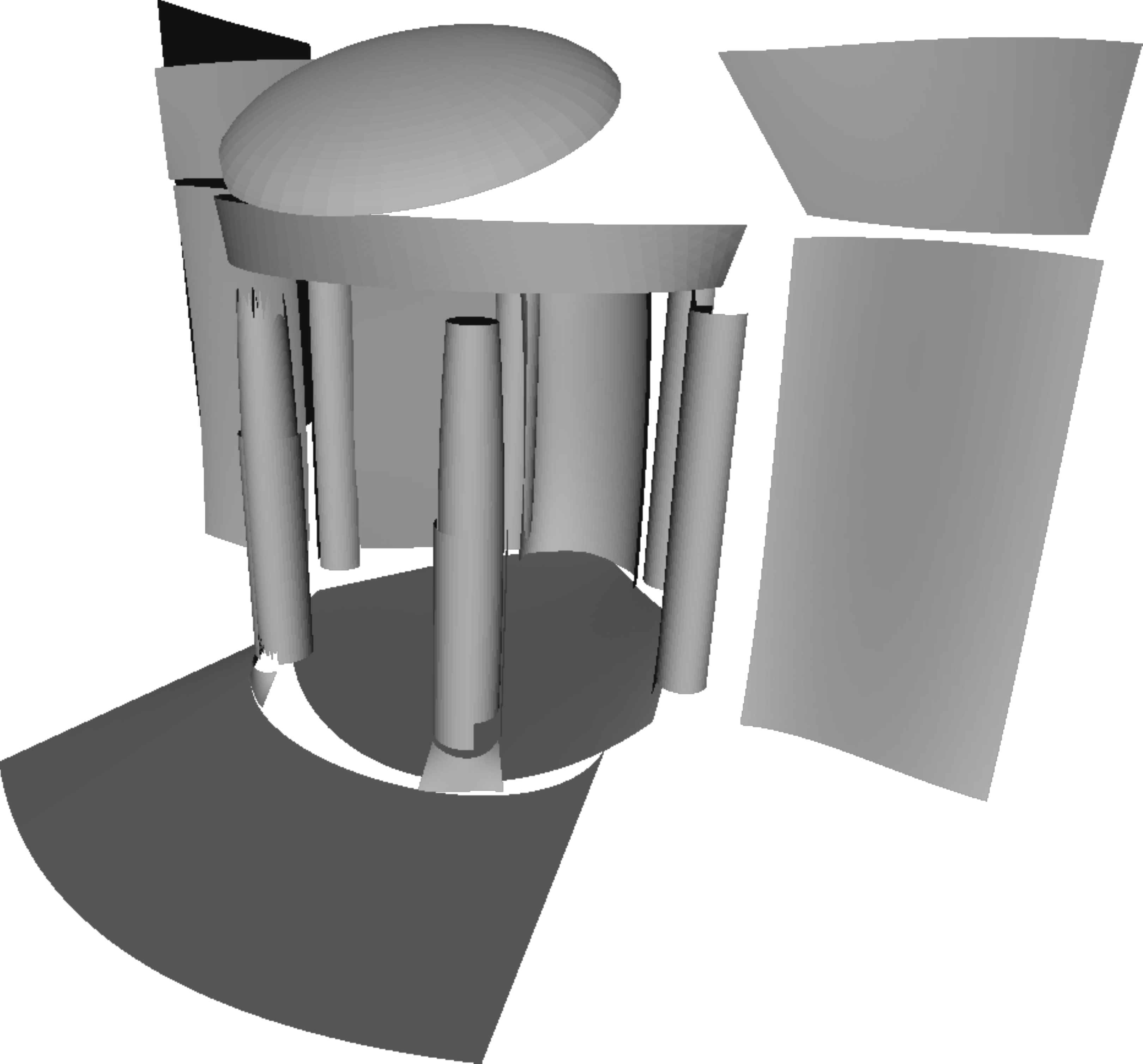} &
	 \includegraphics[width=0.35\textwidth]{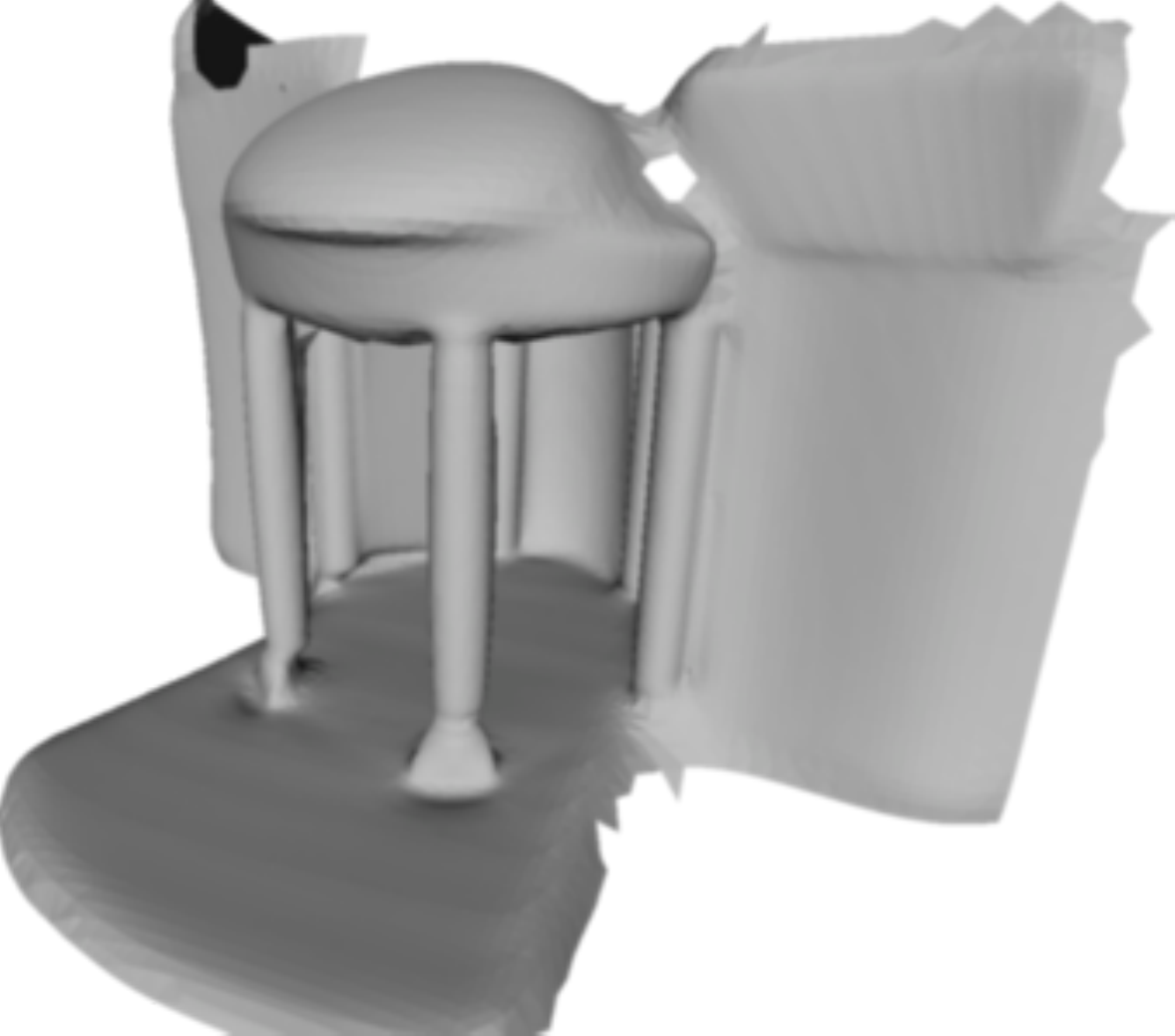}
	 \end{tabular}
	 \caption{After iteratively selecting points in the original cloud, system produces a collection of primitives which form the surface of the required object. The left image shows an example of the extracted geometry, and the right image shows the mesh after it has been closed using Poisson reconstruction.}
	 \label{fig:exampleoutput}
\end{figure}
The system assumes the existence of a point cloud with a collection of calibrated images. The user is presented with two panes, one which displays an interface with the image and allows a frame number to be selected, and the other containing the point cloud and tools to view the cloud from an arbitrary pose and scale. Sketches can be drawn on the image pane with colour selection representing curves which correspond across each of the views.

A schematic diagram for the system is presented in Figure \ref{fig:scheme}. The point clouds are generated through image based reconstruction, using the well known and freely available software VisualSFM~\cite{vsfm1,pmvs}. The rest of this paper describes the interactive modelling, which makes up the principle contribution. The meshes are finalised with a Poisson reconstruction~\cite{kazhdan2006poisson} and then textured.

The system is a free form, colour coded, tool for highlighting corresponding trajectories. The trajectories enable a sub-selection of the point cloud to be extracted for more advanced fitting. The types of surface that can be fit are first classified into `whole surface fitting' and `combining curves'. The former reduces the point cloud to a single point sub-cloud, from which a surface derived from a quadric is fitted, either an Ellipsoid or Cylinder. The latter takes two point sub-clouds representing two curves in three dimensions and combines them through perpendicular extrusion or interpolation. The categories are labelled below,  
\begin{enumerate}
	\item \textbf{Whole surface fitting}. A single point cloud is derived from the user sketches, from which the algorithm fits, either,
	\begin{enumerate}
		\item an \textbf{Ellipsoid},
		\item or a \textbf{Cylinder}.
	\end{enumerate}
	The method for fitting a quadric is described in Section \ref{sec:quadric}, with a discussion on how to obtain the primitive surfaces from the parameters.
	\vspace{0.3cm}
	\item \textbf{Combining curves}. A pair of point clouds are derived from the user sketches, representing a pair of curves in 3D. The curves are then combined to form a surface through;
	\begin{enumerate}
		\item \textbf{Perpendicular extrusion}, one of the curves is copied along the other to form the vertices of a surface;
		\item or, \textbf{Interpolation}, where the curves form the ends of the surface and intermediate vertices are interpolated.
	\end{enumerate}
	The latent variable model for the curves is described in Section \ref{sec:lvm}, from which the surfaces defined above are created.
\end{enumerate}
\begin{figure}[t!]
	\centering
	 \includegraphics[width=0.8\textwidth]{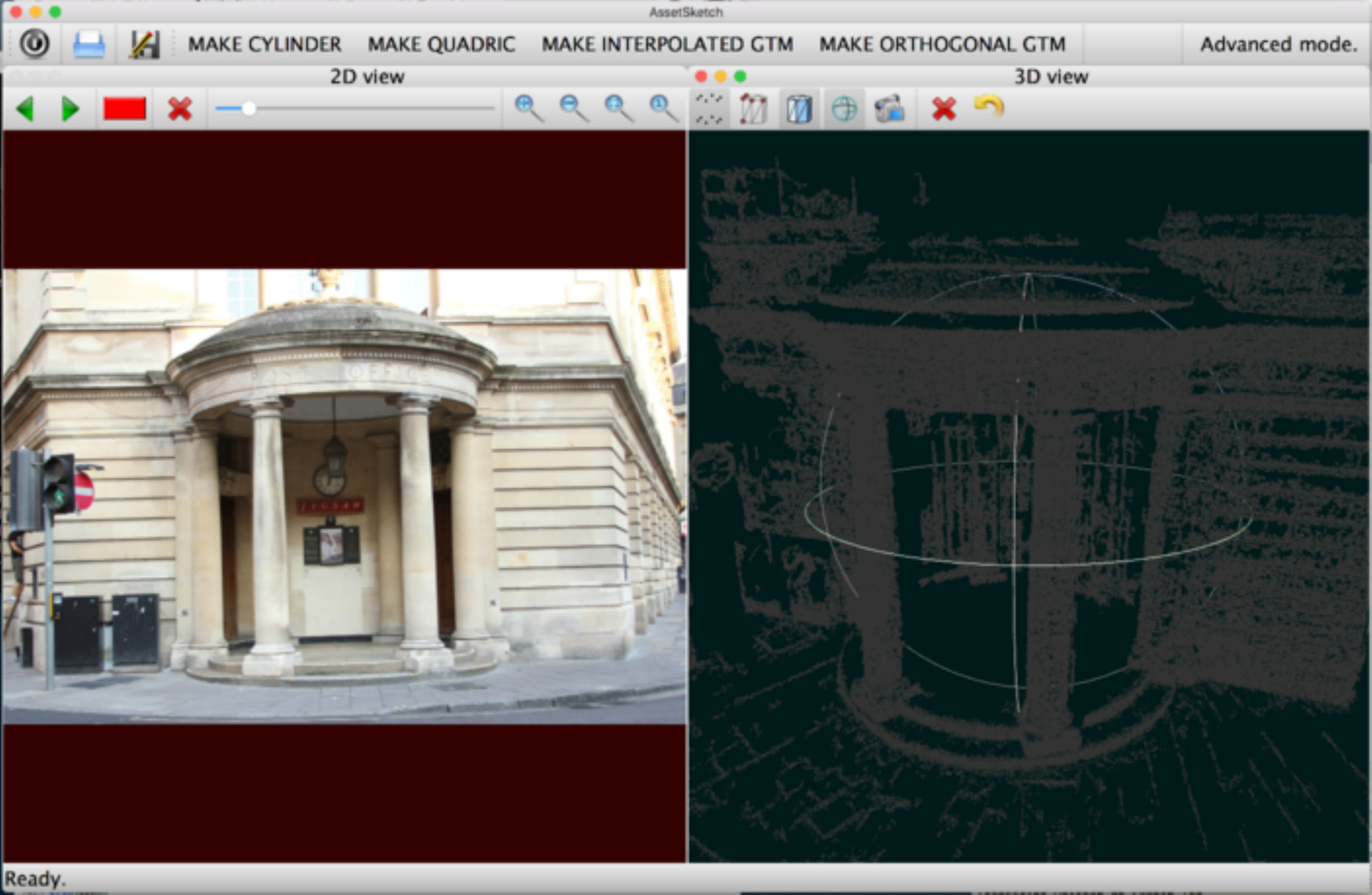}
	 \caption{The graphical user interface.}
	 \label{fig:systemdiagram}
\end{figure}

The graphical user interface is displayed in Figure \ref{fig:systemdiagram}. The user is presented with a panel for sketching on camera views on the left, and with the point cloud on the right. Some examples of fitting primitives to the point cloud using the interface are given in Figure \ref{fig:examplesketchinput}. The top row shows an ellipsoid being fit to the data from two curves on two views, and the bottom row shows a perpendicular extrusion. The process is refined over a number of iterations to produce an output such as on the left of Figure \ref{fig:exampleoutput}. The surface is saved as a collection of meshes which are then processed by a Poisson surface reconstruction algorithm, in order to fill the gaps between the surfaces; displayed on the right hand side of Figure \ref{fig:exampleoutput}. 

The method of free-form point selection is first described in Section \ref{sec:pointselection}, before the various fitting algorithms. The curves are modelled using a latent variable model, this is described in Section \ref{sec:lvm} along with the methods for reconstructing surfaces from curves. The ellipsoid and cylinder fitting algorithms are described in Section \ref{sec:quadric}.

\section{Free-Form Point Selection} \label{sec:pointselection}
The problem of selecting points is difficult if both the input and the point cloud is contains noise. It is natural to appeal to Bayesian statistics, since it is a well established framework for these kinds of problem. Although there are many different types of noise, it is assumed that the user-input and point cloud exhibit additive Gaussian noise, and the corresponding points that were selected in the cloud are extracted by thresholding the probability distribution.

The interface takes a pen width and colour, and allows a user to draw a profile across the image. By marking the image the user is supplying the likelihood that a pixel belongs to a group of 3D points that are of interest in the point cloud, either because the pixels represent the boundary between surfaces or fall within the interior of the surface. It is shown how to combine these distributions in order to derive a probability that each point in the cloud belongs to a sketch. A threshold can then be applied to the density in order to extract points of interest. 

Let the collection of all images be denoted $F = \{f_1, \hdots, f_N\}$. The subset of images that have been marked are denoted $U$, so that $i \in U$ refers to image $f_i$. Typically only a few images are required for marking and so the size of $U$ is small compared to $F$, particularly in the case of a video.

Points along the user sketch are denoted $\bm{\mu}_{ij}$; this is the $j^{th}$ point in image $f_i$. Each point has an associated covariance $\Sigma_{ij} = \sigma^2_{ij}I$; the $\sigma_{ij}$ represents the uncertainty in the estimated edge location and are chosen to be the brush width in the user interface. If we now consider an arbitrary pixel $\bm{y}\in \mathbb{N}^2$, in an image. the likelihood that this pixel belongs to the mark drawn by the user is modelled as a Gaussian mixture,
\begin{align}
	p(\bm{y} | f_i) = \frac{1}{K_i} \sum_{j=1}^{K_i} \mathcal{N}(\bm{y}|\bm{\mu}_{ij}, \Sigma_{ij})
\end{align}

Now suppose that we have an arbitrary 3D point $\bm{z} \in \mathbb{R}^3$, and a collection of cameras $\Pi = \{\pi_1, \hdots, \pi_{|U|}\}$,
\begin{align}
	\pi_i : \mathbb{R}^3 \rightarrow& \mathbb{R}^2 \\
	\bm{z} \mapsto& \frac{\left[ \begin{array}{c c c}1 & 0 & 0 \\ 0 &1 & 0\end{array}\right] P_i\left[ \begin{array}{c}\bm{z}\\1\end{array}\right]}{(P_i\left[ \begin{array}{c}\bm{z}\\1\end{array}\right])_4} \label{eq:projection}.
\end{align} 
where $P_i \in \mathbb{R}^{3 \times 4}$ is the $i^{th}$ camera matrix.

We wish to find the overall probability of the point in the cloud $\bm{z}_k$, where $0 < k < N$. In order to do it we assume that the sketches are conditionally independent given the projection functions and the 3D point, in which case the distribution can be written as follows,
\begin{align}
	p( \bm{z}_k | \Pi ) = \frac{1}{\mathcal{C}}\prod_{i = 1}^{|U|} p(\pi_i(\bm{z}_k) | f_i)
\end{align}

The normalising factor $\mathcal{C}$ is then the sum over each point in the cloud,
\begin{align}
	\mathcal{C} = \sum_{k=1}^N\left(\prod_{i = 1}^{|U|} p(\pi_i(\bm{z}_k) | f_i) \right)
\end{align}

We use this probability to identify 3D points among the million or so points in the cloud. Every point in the cloud is assigned a probability, many of them very low. We simply keep all of the points in the cloud which have probability larger than the mean probability. 

It would be possible to include further interaction to identify the surface normals. This is not necessary for this method since the positions of the cameras are known. The interior of the surface is set so that all of the normals are pointing towards a candidate camera.

\section{Fitting a Quadric} \label{sec:quadric}
The first category of point extraction is termed `whole surface fitting', in which the user selects a region which contains points that belong to a whole surface. Either an ellipsoid or a cylinder is then fitted. Both of these surface types can be neatly represented as a quadric. A full and detailed explanation of the method can be found in \cite{bealetal}. The salient points are described in this section, along with the method that is used for trimming the final surfaces.

Quadrics are implicit functions, specifically, solutions of the equation,
\begin{align}
	\bm{z}^T A \bm{z} + \bm{b}^T\bm{z} + c = 0 \label{eq:quadric}
\end{align}
with $A \in \mathbb{R}^{3\times 3}$ matrix which is symmetric positive definite, $\bm{b} \in \mathbb{R}^3$ a vector and $c \in \mathbb{R}$ a scalar. 

In three dimensions $A$ and $\bm{b}$ can be written as follows,
\begin{align}
	A = \left[ \begin{array}{c c c} a_{11} & a_{12} & a_{13} \\
			a_{21} & a_{22} & a_{23} \\
			a_{31} & a_{32} & a_{33}  \end{array} \right]
	\bm{b} = \left[ \begin{array}{c} b_1 \\ b_2 \\ b_3 \end{array} \right]
\end{align}

We let,
\begin{align}
	\bm{\psi} = [a_{11}, a_{22},a_{33},a_{12},a_{13},a_{23},b_1,b_2,b_3,c]^T
\end{align}

Suppose that the matrix of data points are denoted $Z = [\bm{z}_1, \hdots, \bm{z}_N]$ with the $k^{th}$ column denoted $\bm{z}_k \in \mathbb{R}^3$, and let the $j^{th}$ element of $\bm{z}_k$ be $z_{jk}$. We create a vector of monomials for each data point as follows,
\begin{eqnarray}
	\bm{x}_i& = [&z_{1i}^2, z_{2i}^2, z_{3i}^2, 2z_{1i}z_{2i},  
				2z_{1i}z_{3i}, 2z_{2i}z_{3i}, \nonumber \\ &&z_{1i}, z_{2i}, z_{3i}, 1]^T
\end{eqnarray}

If we define a monomial data matrix $X = [\bm{x}_1, \hdots, \bm{x}_N]$, the maximum a-priori solution for the parameters is,
\begin{align}
	\bm{\psi} = \left( \sigma^2 X^TX + \left[ \begin{array}{cc} J & \bm{0} \\ \bm{0} & \bm{0} \end{array} \right] \right)^{-1} \left[ \begin{array}{c} \bm{1} \\ \bm{0} \end{array}\right]
\end{align}

Where $\sigma$ is a prior parameter which controls how close to a sphere the quadric is. The smaller it is, the more likely that the quadric will take the shape of a sphere. This method is useful since it prevents the quadric from forming a hyperboloid which is made of two disconnected surfaces. Usually when fitting primitive geometry we are looking for a single closed surface, and the best fit to be a contiguous collection of points.

In practise, the surfaces must be trimmed so that they do not go too far beyond the point cloud. This is done by considering the eigen-decomposition of the matrix $A$. Considering the expansion,
\begin{align}
	& (\bm{x} - \bm{\mu})^T A (\bm{x} - \bm{\mu}) = \tau \label{eq:quadfactors} \\
	\iff& \bm{x}^T A \bm{x} - 2 \bm{\mu}^T A \bm{x} + \bm{\mu}^T A \bm{\mu}
\end{align}
and taking $\bm{b} = - 2A\bm{\mu}$ and $c = \bm{\mu}^T A \bm{\mu}$, it can be seen that Equations \ref{eq:quadfactors} and \ref{eq:quadric} are equivalent. Removing the mean and projecting the points onto the principal vectors provides a way to determine how much of the quadric is covered by the points. To trim the final surface we simply remove vertices and faces which are outside the maximum and minimum projected values.

\begin{figure}[t!]
\begin{center}
\begin{tabular}{>{\centering\arraybackslash}p{0.3\linewidth} >{\centering\arraybackslash}p{0.3\linewidth} >{\centering\arraybackslash}p{0.3\linewidth}}
\includegraphics[width=0.7\linewidth]{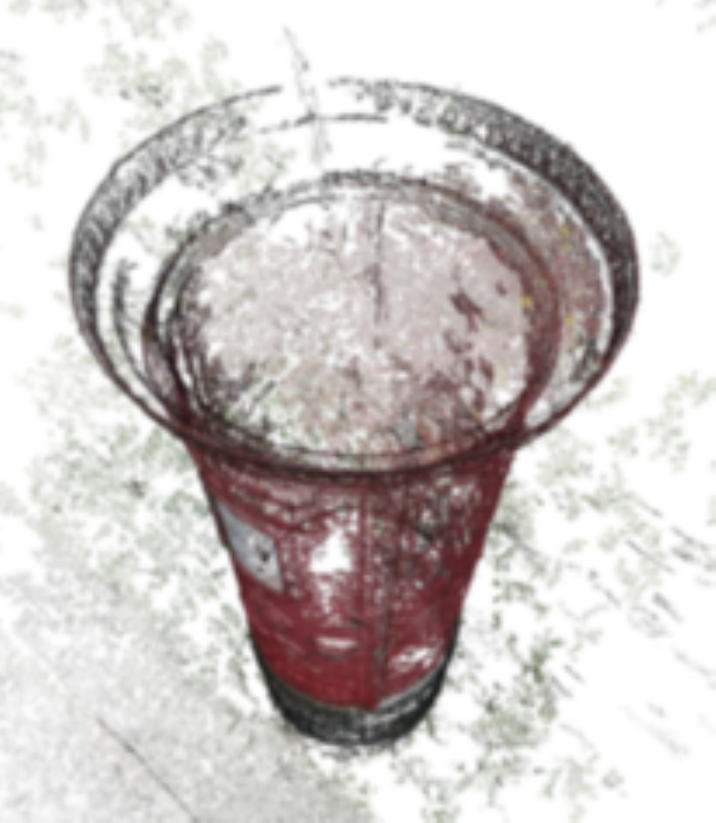}
& \includegraphics[width=0.7\linewidth]{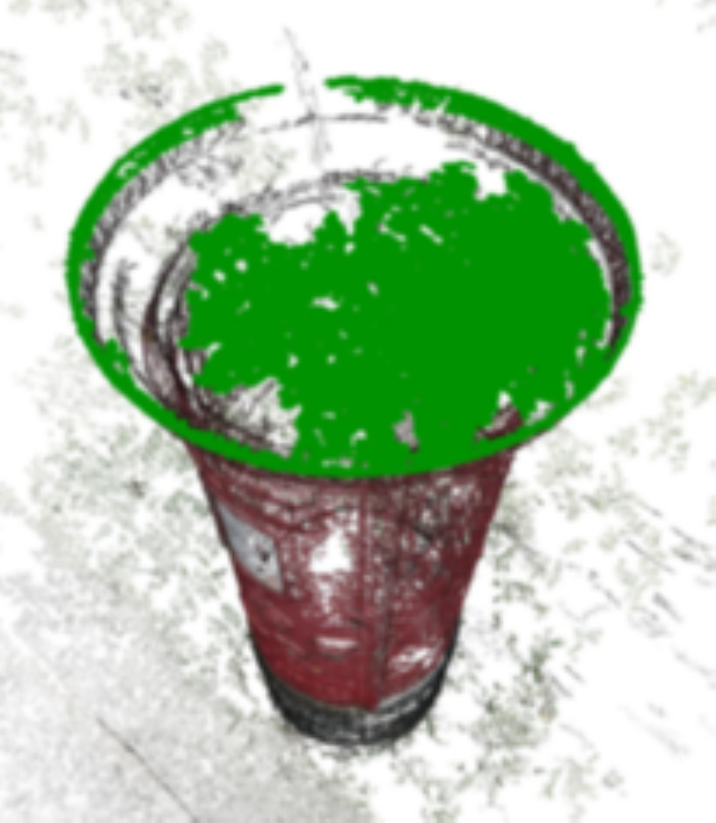}
& \includegraphics[width=0.7\linewidth]{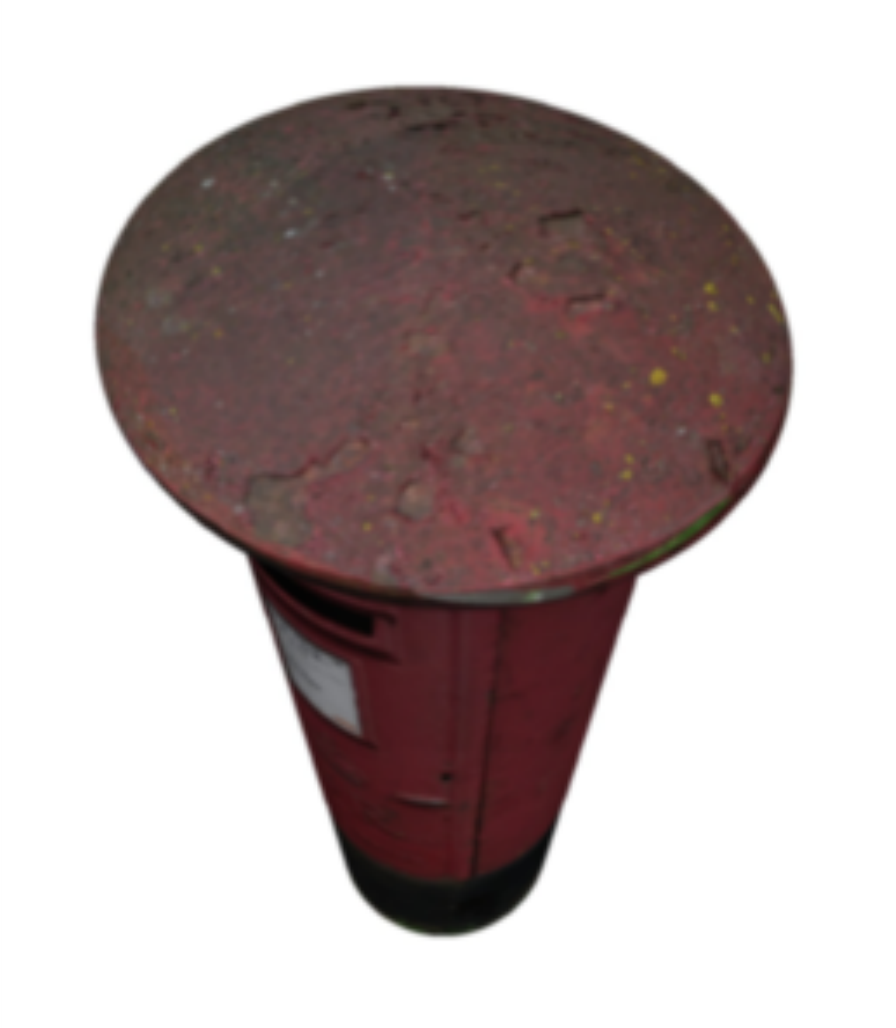}\\[5pt]
\textit{(a)~Point cloud} & \textit{(b)~Points selected by user sketches} & \textit{(c)~Recovered surface} 
\end{tabular}
\caption{An example of fitting a quadric to a point cloud. (a)~The point cloud. (b)~Points selected by user sketches are shown in green. (c)~The final quadric surface fit to the point cloud. The surface extends to the extremal points selected by the users sketches or curves.}
\label{fig:fitting_post_box_roof}
\end{center}
\end{figure}

Figure \ref{fig:fitting_post_box_roof} shows an example of fitting a quadric surface to a set of 3D points selected from the point cloud.

\section{Fitting a Latent Variable Model} \label{sec:lvm}
The second categories of point extraction is termed `combining curves'. In this mode the user selects a pair of curves in the interface which are then combined either through interpolation of perpendicular extrusion in order to create the surface. This section describes the method used for fitting a parametric function to the curves and then combining them in to a single surface.

One of the principal difficulties in fitting a parametric model to a point cloud is that the points are noisy and unordered. We suppose that the points relate to the output of a continuous function, of a single variable. In other words given a value $t$ we would like to be able to compute an arbitrary point on the curve. If we know how the value of $t$ relates to the output, the problem is one of standard regression, if we do not, the $t$ values are unknown or hidden, and are termed `Latent Variables'. A number of different methods for extracting latent variables and fitting parameters are documented in \cite{lawrence2004gaussian}. Following the work of \cite{bishop1998gtm}, we make use of the Generative Topographic Map (GTM), since it is flexible to a range of different types of function, requiring only a finite set of basis vectors.

There are several different types of curve, such as B-Splines, Bezier curves, or linear combinations of sinusoids or exponential kernels. A multidimensional polynomial is chosen since it is simple and portable. Letting $\bm{\psi}_d \in \mathbb{R}^L$ denote the basis vectors for $ 0 < d < D$, and the matrix $\Phi = [\bm{\phi}_0, \hdots, \bm{\phi}_D]$. For a polynomial basis this is written as follows,
\begin{align}
	\Phi = \left[ \begin{array}{c c c c c} 1 & 1 & 1 & \hdots & 1 \\ 0 & 1 & 2 & \hdots & D \\ \vdots & \vdots & \vdots & & \vdots \\ 0^L & 1^L & 2^L & \hdots & D^L  \end{array} \right] \\
\end{align}

\begin{figure}[t]
\centering
\includegraphics[width=0.4\textwidth]{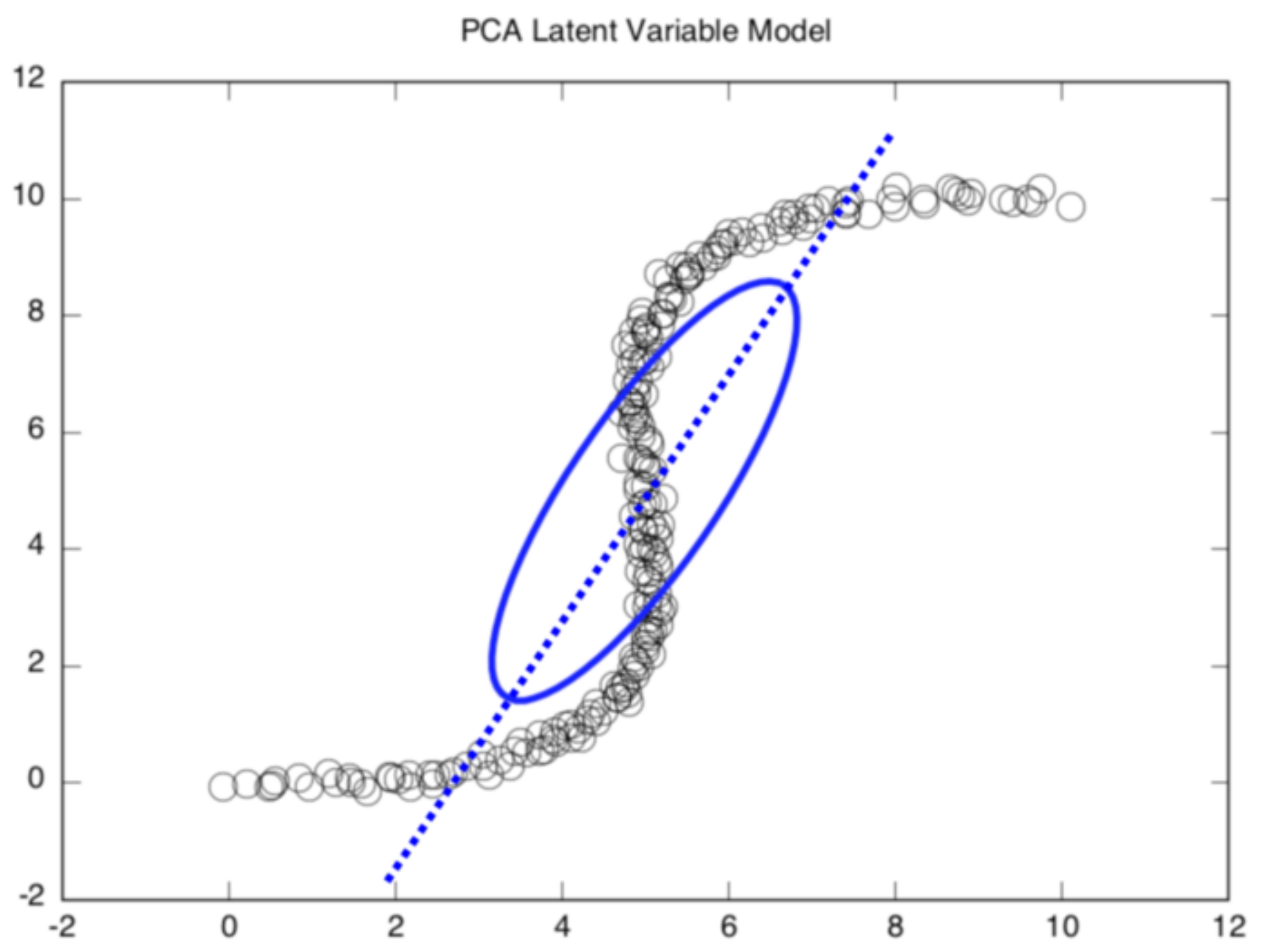} 
\includegraphics[width=0.4\textwidth]{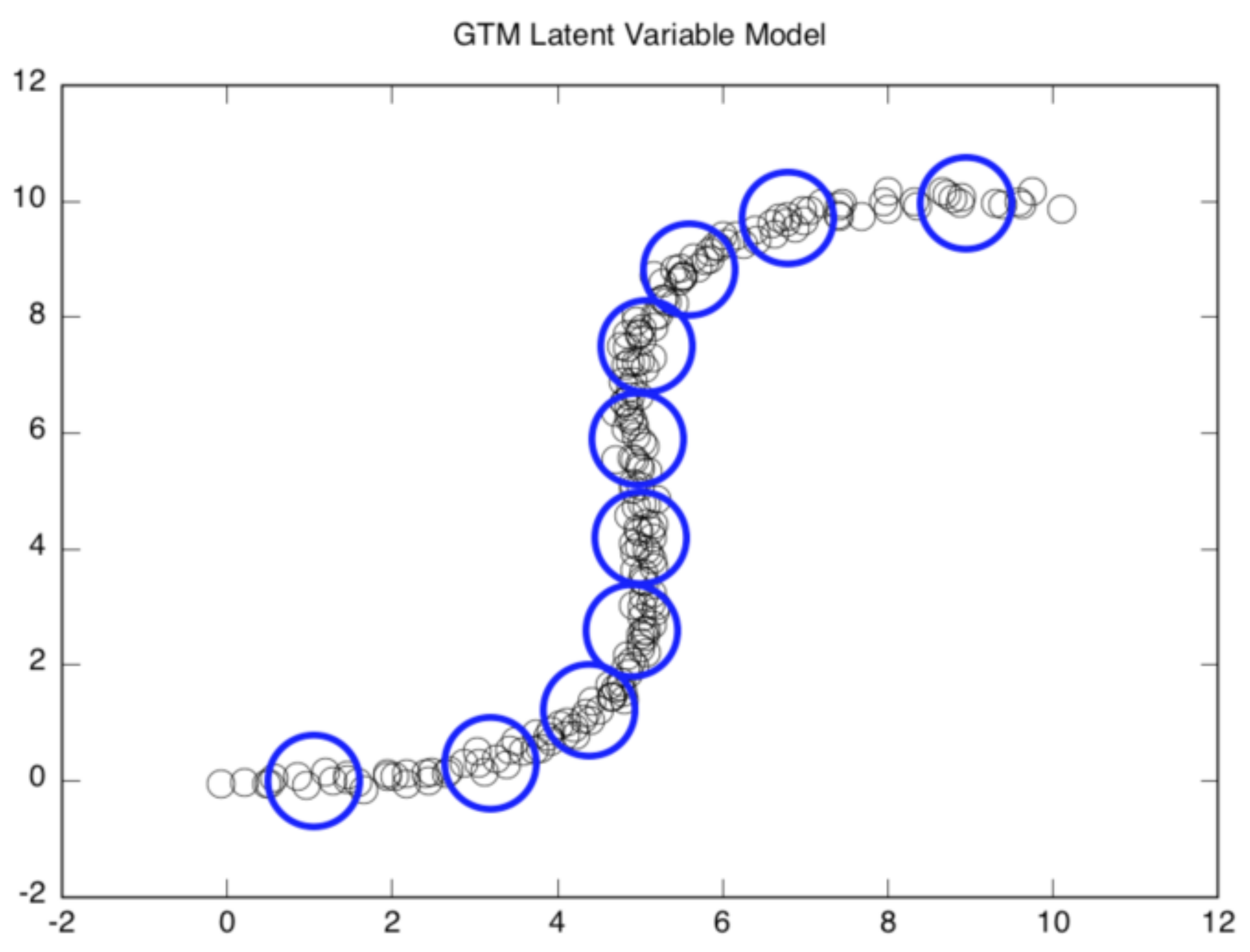}
\caption{An example of multivariate polynomial regression using both Principle Component Analysis (left) and a Generative Topographic Map (right). The black points are the noisy point cloud in 2D, the blue ellipses represent Gaussians. }
\label{fig:regress}
\end{figure}

Given a weight matrix $W \in \mathbb{R}^{3\times L}$, the product $W\Psi$ is then a collection of polynomials evaluated at each of the points $0, \hdots, D$; with coefficients in $W$. The value of $D$ can be chosen by the user, but is generally fixed at about $50$. The values of $W\Psi$ are mapped in to the 3D point cloud space, defining the equidistant centres of a Gaussian mixture along the polynomial. The model is best represented by Figure \ref{fig:regress} (right) where the black points are unordered 2D data distributed along the curve, and the curve itself is shown in blue.

The constrained Gaussian mixture model, for the collection of points $ Z = [Z_1, \hdots, Z_N]$ is given by,
\begin{align}
	p(Z | \Phi, W) = \prod_{k=1}^{N} \left( \sum_{j=1}^{D} \omega_k\mathcal{N}(Z | W\bm{\phi}_j, \sigma^2I) \right) \label{eq:gtmmixture}
\end{align}

The mixture is solved for the parameters $W$ and $\sigma$ by finding the derivatives of the log-likelihood of Equation \ref{eq:gtmmixture}. The solution is a fixed point iteration, which can be formulated as an Expectation-Maximisation algorithm as follows. We first compute the expected probability of a point given the parameters and the data,
\begin{align}
	R_{ij} = \frac{\omega_i \mathcal{N}(Z_j | W\bm{\phi}_i. \sigma^2I)}{\sum_{m=1}^D \omega_m \mathcal{N}(Z_j | W\bm{\phi}_m, \sigma^2I)}
\end{align}

The maximisation step finds the maximum likelihood values for the parameters given the expected probabilities,
\begin{align}
	W^{T}_{new} =& (\Phi G \Phi^T)^{-1} \Phi R Z^T \\
	\sigma^2 =& \sum_{i=1}^D \sum_{j=1}^K R_{ij} || W_{new}\bm{\phi} - Z_j||^2
\end{align}
where G is a diagonal matrix with elements $ G_{ii} = \sum_{n}R_{in} $.

Since this procedure is iterative it requires an initialisation of the variables $W$ and $\sigma$. This is done using principle component analysis. Consider a line defined by the function $f(t) = \bm{n}t + \bm{a}$, and a Gaussian which has been fit to the data defined by by parameters $\bm{\mu}$ and $\Sigma$. We would like the line to pass through the mean of the data and along the largest principle component of $\Sigma$. This is shown graphically on the left of Figure \ref{fig:regress}, the unordered data is represented by the black circles and we would like to fit the line represented by the blue dashes. The parametrisation of $f$ makes makes this easy: set $\bm{a}=\bm{\mu}$ and let $\bm{n}$ be the vector corresponding to the largest eigenvalue of $\Sigma$. In the notation of the above the weight matrix is then taken to be $W = [\bm{a}, \bm{n}, \bm{0}, \hdots, \bm{0}]$, and the variance is initialised as the smallest eigenvalue of the covariance matrix.

Generally speaking, the ends of the curve do not need to be computed since the algorithm converges to fit the points precisely. It is possible that the shape of the curve causes the Gaussians to be centres beyond the points. In this case the GTM is trimmed by removing components at either end which have a value of $\frac{1}{K} \sum_{k=1}^K R_{ij}$ that falls below a threshold, or in other words, Gaussian components are removed from the ends which do not have any points associated with them. Doing this keeps the resulting curve tight to the original data, even after convergence.

The final surface is built from the curves. Either the surface is an interpolation of two drawn edge curves, or one curve is extrapolated in the direction of a perpendicular second curve. In practise, drawing a profile and extrapolating it along an edge gives very good results; a similar approach was taken in 3-sweep~\cite{chen2008sketching}.

Letting $\bm{q}_i, \bm{p}_i \in \mathbb{R}^3$ denote two extracted curves, and $\bm{s}_{kj}$ be a set of vertices on the desired surface, for $0 < i, j, k \leq D$. We first reorder the curves so that they are oriented in the same direction; if $|| \bm{q}_0 - \bm{p}_D|| < || \bm{q}_0 - \bm{p}_0 || $ then the curve $(\bm{p}_j)_{j=1}^{D}$ is reversed. The interpolation is then performed using the following equation,
\begin{align}
	\bm{s}_{kj} = \left( \frac{j}{D} \right )\bm{q}_k + \left( 1 - \frac{j}{D} \right ) \bm{p}_k
\end{align}

If the curves are perpendicular then the curves are reordered in the same way as above. One side of the surface is initialised with a trajectory and then copied and translated it according to the velocities of the second. This method avoids the requirement of computing curve intersections, which may be difficult if the data is noisy. The surface vertices are as follows,
\begin{align}
	\bm{s}_{kj} = \left \{ \begin{array}{r l} \bm{q}_k & : j=0 \\ \bm{q}_k + \bm{p}_j - \bm{p}_{j-1} & :j>0 \end{array} \right .
\end{align}

\begin{figure}[t!]
\centering
\setlength{\tabcolsep}{5pt}
\begin{tabular}{>{\centering\arraybackslash}p{0.3\linewidth} >{\centering\arraybackslash}p{0.3\linewidth} >{\centering\arraybackslash}p{0.3\linewidth}}
\includegraphics[width=0.7\linewidth]{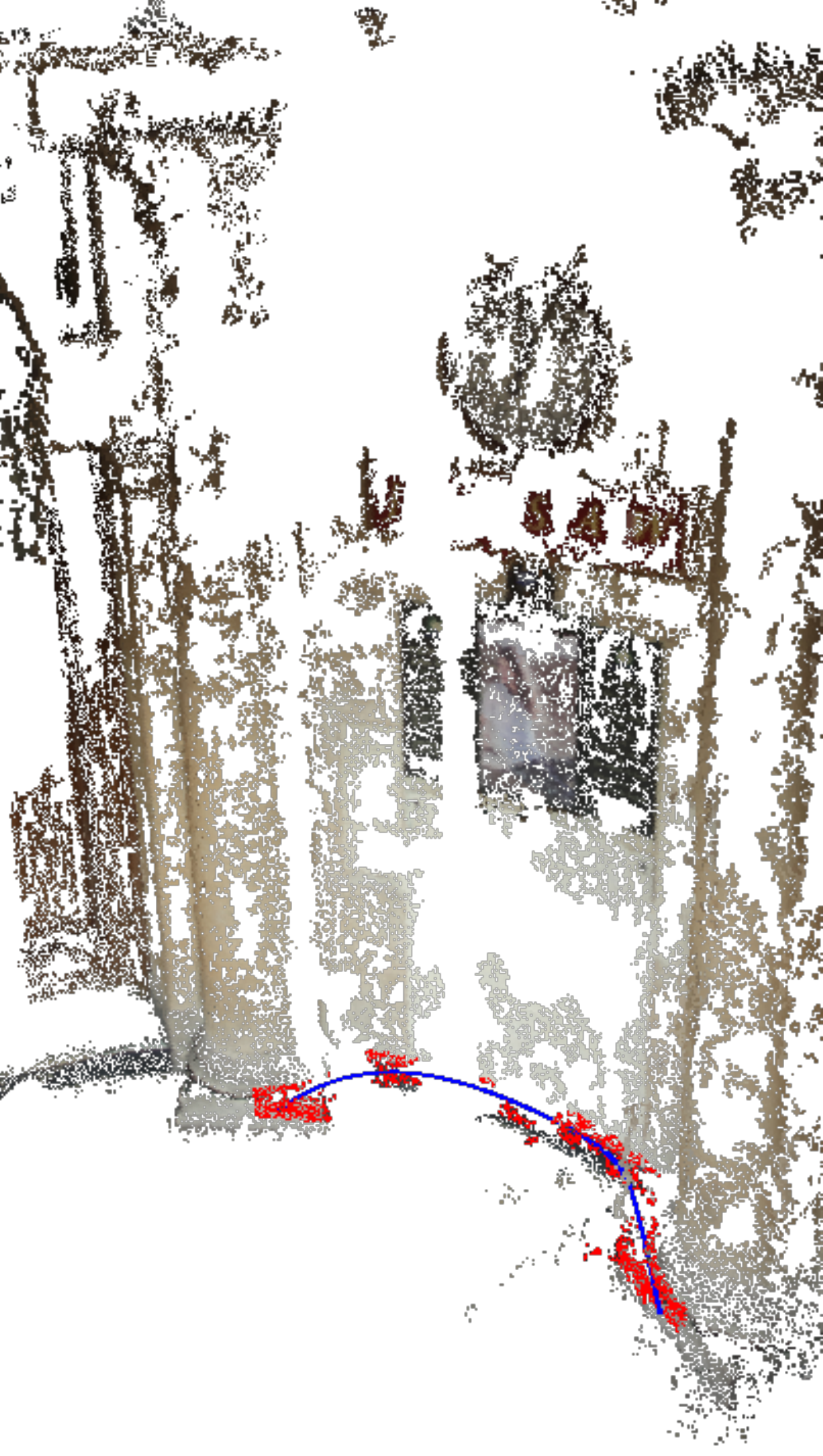} &
\includegraphics[width=0.7\linewidth]{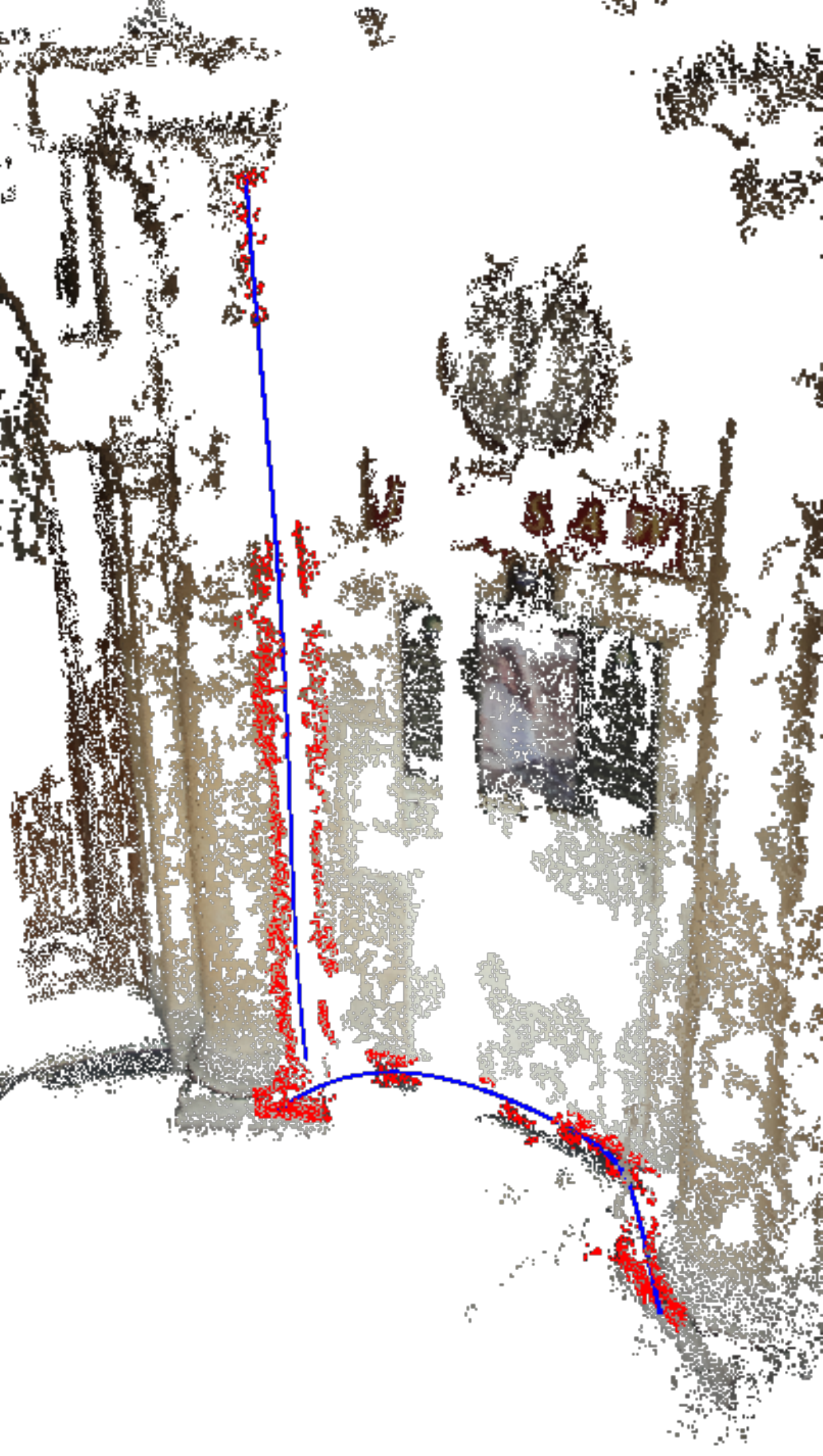} &
\includegraphics[width=0.7\linewidth]{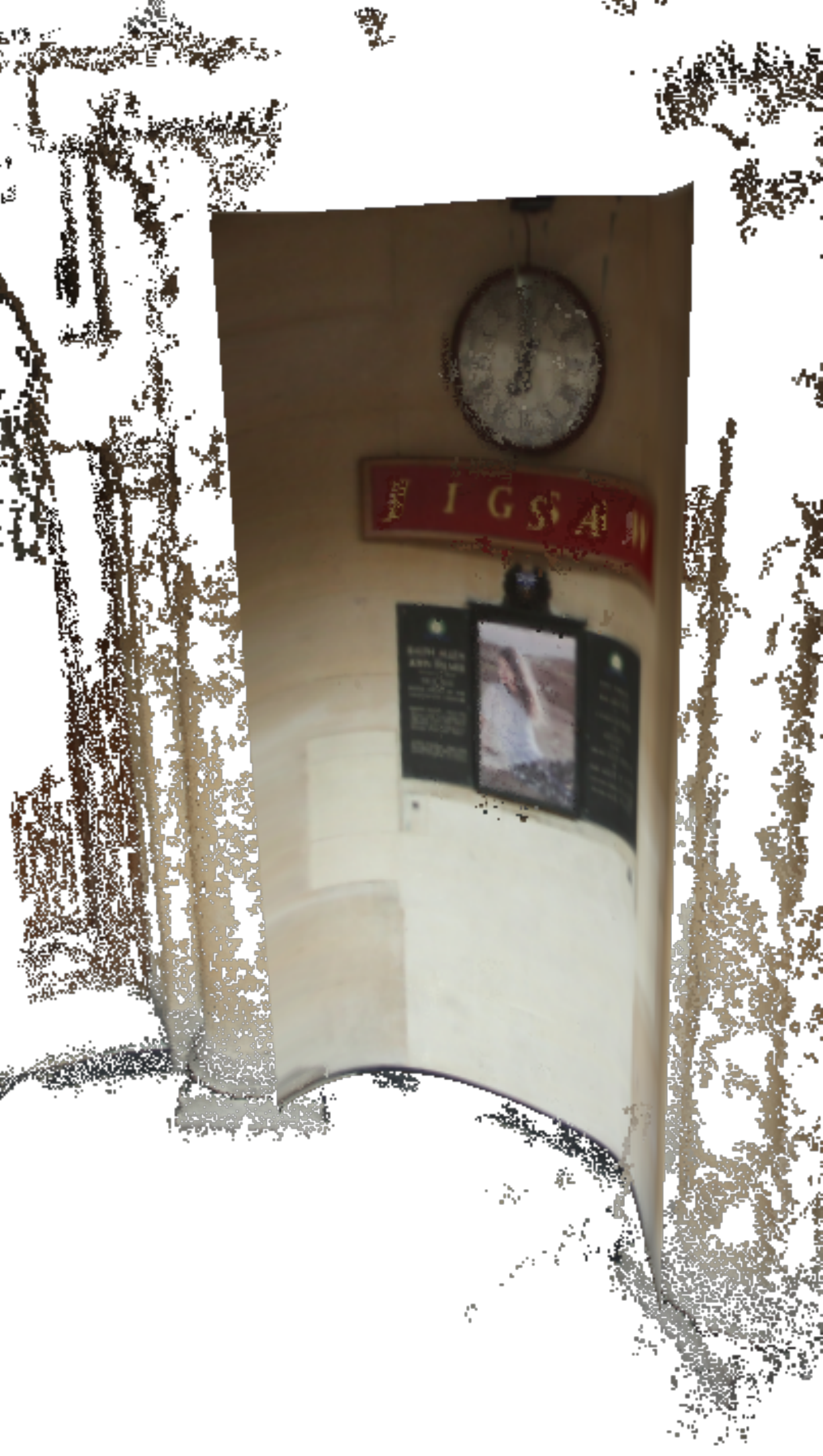} \\
\textit{(a)~First edge} & \textit{(b)~Second edge} & \textit{(c)~Final surface}
\end{tabular}
\caption{An example of fitting a perpendicular GTM surface from a pair of curves. A single collection of sketches define each of the edges, and the first edge is extruded along the second edge to create the final surface.}
\label{fig:fitting_surface_to_curve}
\end{figure}

An example of fitting a perpendicular GTM surface is shown in Figure \ref{fig:fitting_surface_to_curve}. 

\section{Finalising the surface}
The system described in previous sections provides a way for a user to fit geometric primitives to a point cloud. It remains to fill in the holes in the surfaces. 

To close the holes between the primitive surfaces the Poisson reconstruction algorithm is used, as described in \cite{kazhdan2006poisson}. An open source implementation is available in the software package MeshLab \cite{cignoni2008meshlab}. Since the output from the software is a collection of meshes with normals no pre-processing is required, and a closed surface is fit to the geometry. The mesh is completed by removing all of the vertices and faces which have long edges. This is possible since the Poisson reconstruction tends to produce large triangles in regions where it is predicting. 

\section{Results}
The results in this section present comparisons against state of the art alternatives for reconstructing surfaces from point clouds. It is difficult to assess the quality of the output without first establishing a high quality ground truth. The problem of overfitting in statistical regression motivates an argument that, even with a good quality ground truth, the quality of a result is difficult to measure, and ultimately lies with the perceiver. Another difficulty in assessing the results is in the human interface. The competing methods require user interaction in order to fit the surfaces. Rather than make arguments based on the effort or complexity of the user input, a single working example is chosen with rational arguments about the limitations of the methods. It is argued that each of the competing softwares are fit for the purpose that they were designed, but fail on the dataset that we have, making the method presented in this paper a reasonable replacement.

The interactive methods that compared against are, VideoTrace \cite{Hengel:2007aa} and Morfit~\cite{yin2014morfit}, both of which are contemporary, and present an alternative approach to surface reconstruction. The comparison is made on a point cloud created from photographs of a building with columns and a non-planar facade. A large portion of the entrance is occluded and there are holes in the point cloud as a result of it. These problems make it very difficult to automatically reconstruct the surface, and a user must provide some idea of where the surfaces should be. This is the example that is used to make the initial comparisons in Sections \ref{sec:morfit} and \ref{sec:videotrace}.

The principal argument for using the method presented in this paper is that a large percentage of man-made objects are composed of a collection primitive surfaces. The final set of results show the performance on a dataset containing 6 examples of buildings and outdoor furniture, the comparisons provide a visual assessment of the improvement beyond the standard method of normal estimation and Poisson reconstruction.

\subsection{Comparison against Morfit} \label{sec:morfit}
The Morfit~\cite{yin2014morfit} software package generates a surface based on the morphological skeleton of the object. The software allows a  user to edit the skeleton, and then sketch extra vertices and normals on a plane which is orthogonal to an arbitrary point on the skeleton, before fitting surface. So long as the skeleton is a good enough prediction the surface will allow a user to specify the surface shape in regions where there are large holes. 

\begin{figure}[h!]
	\centering
	\def\arraystretch{2}
	\begin{tabular}{c c}
		Point cloud & Skeleton \\
		\includegraphics[width=0.4\textwidth]{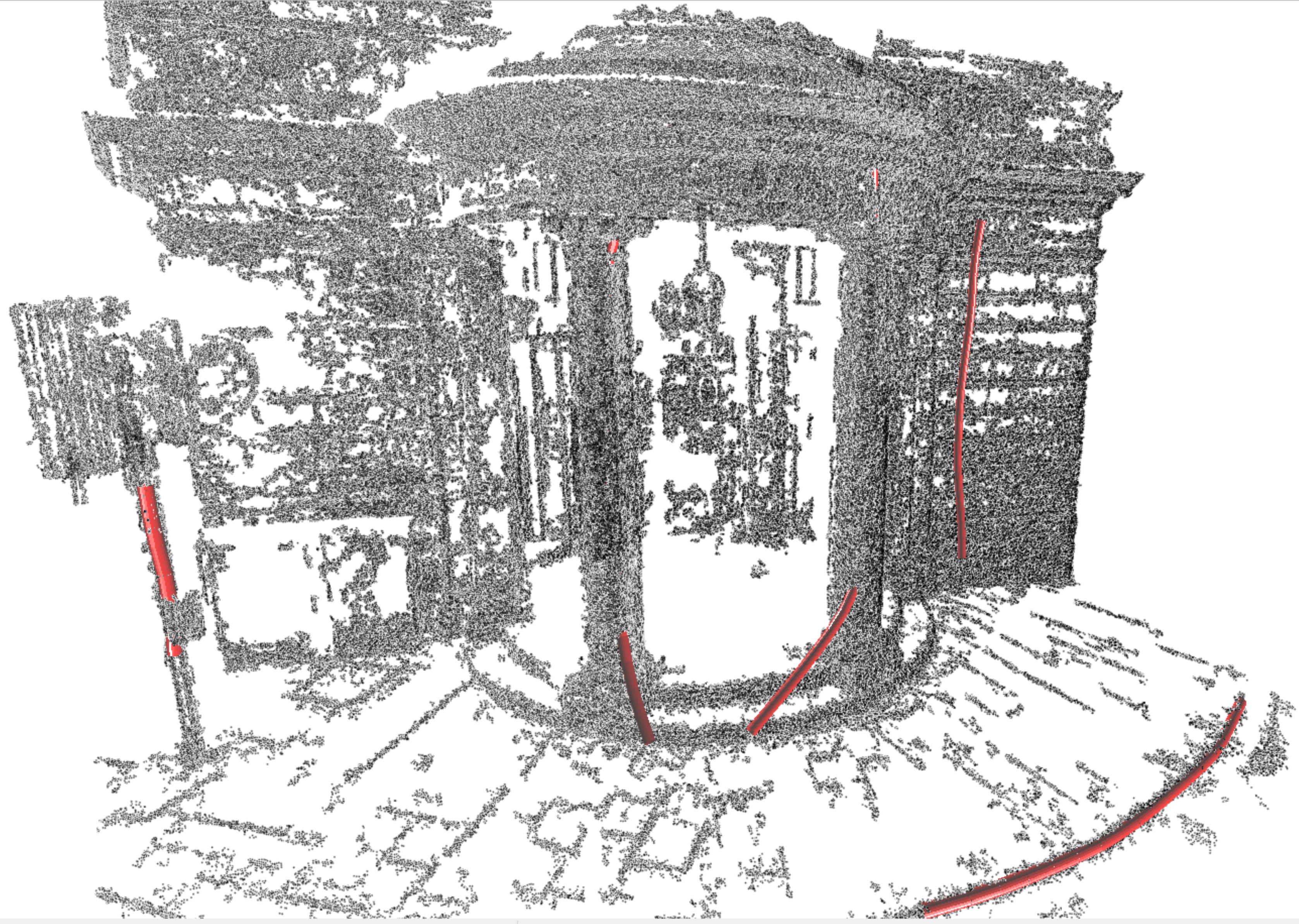} &
		\includegraphics[width=0.4\textwidth]{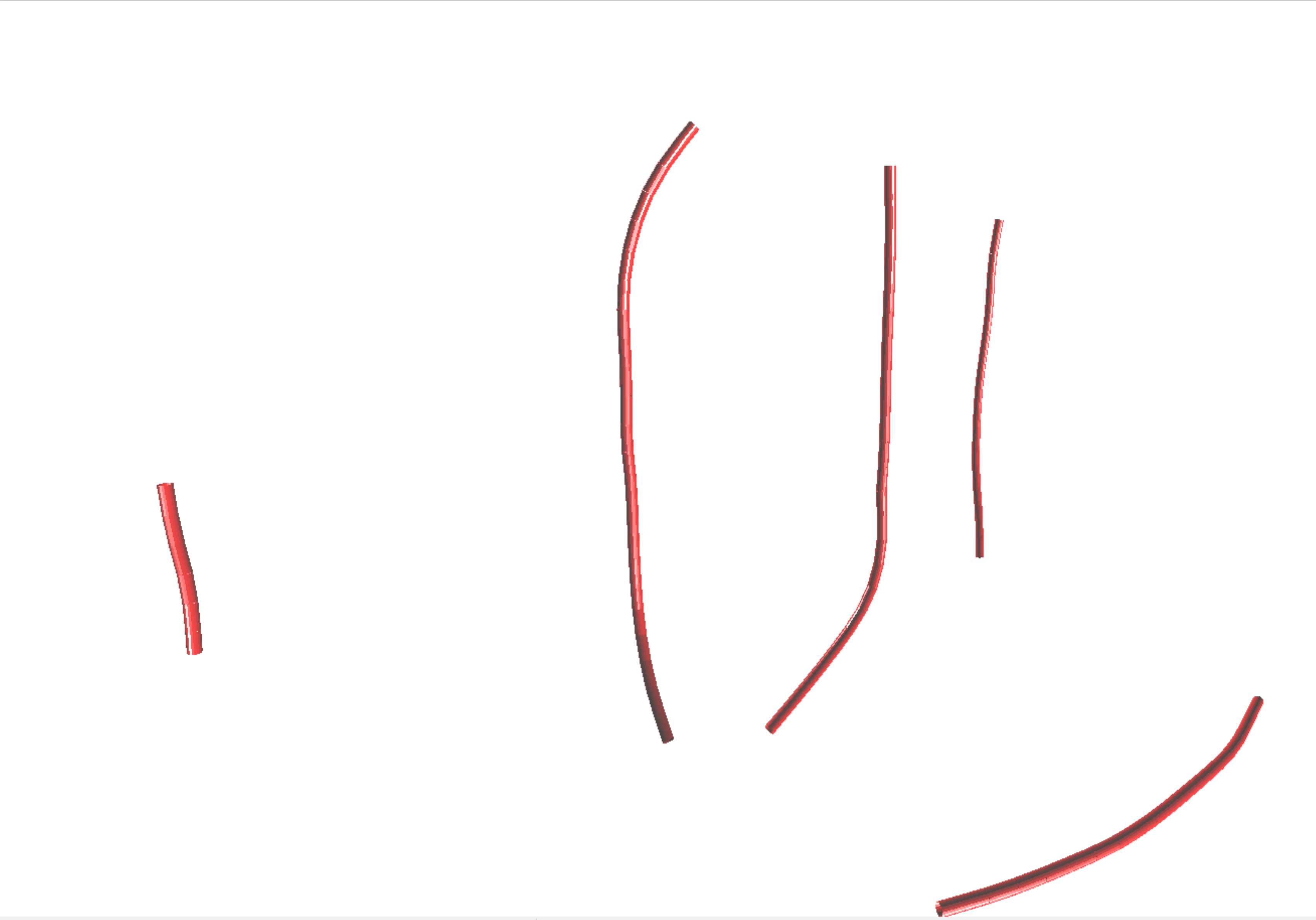} 
	\end{tabular}
	\caption{An example of a skeleton fitted to point cloud of a building with a non-planar and partially occluded facade.}
	\label{fig:morphskel}
\end{figure}

The principle limitation of Morfit is that it requires a morphological skeleton in order to compute the surface. Consider the point cloud shown in Figure \ref{fig:morphskel}. The building has 6 columns arranged in a cylinder, the occluded entrance is cylindrical and each of the walls are planar. The right hand side of Figure \ref{fig:morphskel} shows the automatically extracted skeleton which correctly identified the two front columns, but was unable to find a skeleton for other features such as the back wall. In fact, a skeleton does exist for this building. It is part of a larger closed structure which is only partially visible. Using the Morfit interface to fix the skeleton is extremely time consuming, and running the sweeper on the resulting branches yields poor results.

The dataset that Morfit was tested on is primarily made up of models which are small and closed, such as the model Gekko featured throughout the paper. The algorithm works well on this data, since the skeleton is much easier to compute and the reconstructed points are evenly distributed along the object.

\subsection{Comparison against VideoTrace} \label{sec:videotrace}
VideoTrace is an interactive software package for reconstructing 3D surfaces from a collection of photographs of an object from multiple views. It first reconstructs a sparse point cloud and then allows a user to draw a polygon or non-uniform rational b-spline in three dimensions using the interface. The curves can be joined to create a surfaces, which can then either be extruded or mirrored two create a more complicated closed surface. 

\begin{figure*}[t!]
	\centering
	\def\arraystretch{2}
	
	\begin{tabular}{c c c}
		\multicolumn{3}{c}{\textbf{VideoTrace}} \\
		View 1 & View 2 & Reconstruction \\
		\includegraphics[width=0.23\textwidth]{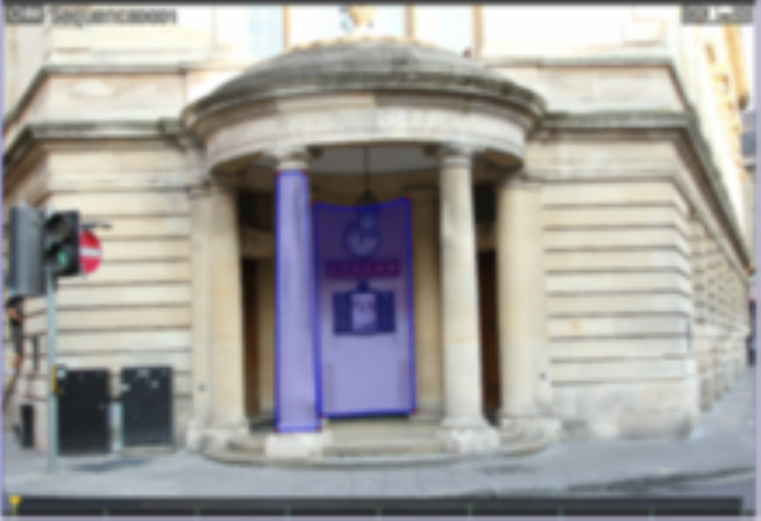} & 
		\includegraphics[width=0.23\textwidth]{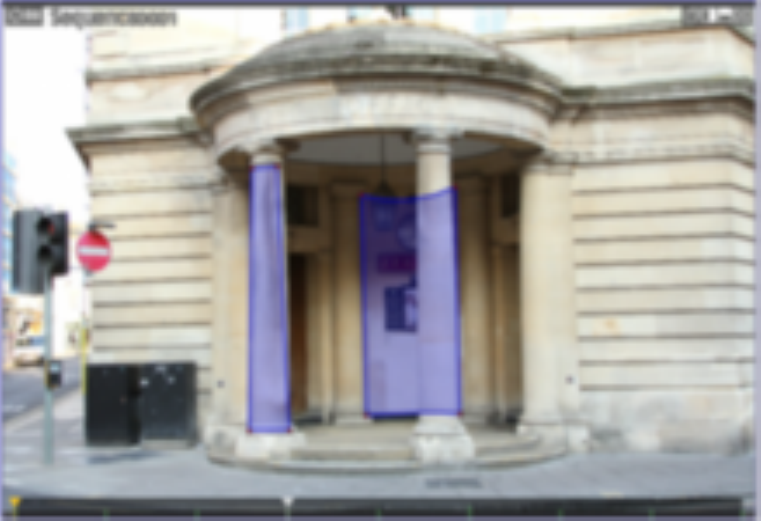} &  
		\includegraphics[width=0.23\textwidth]{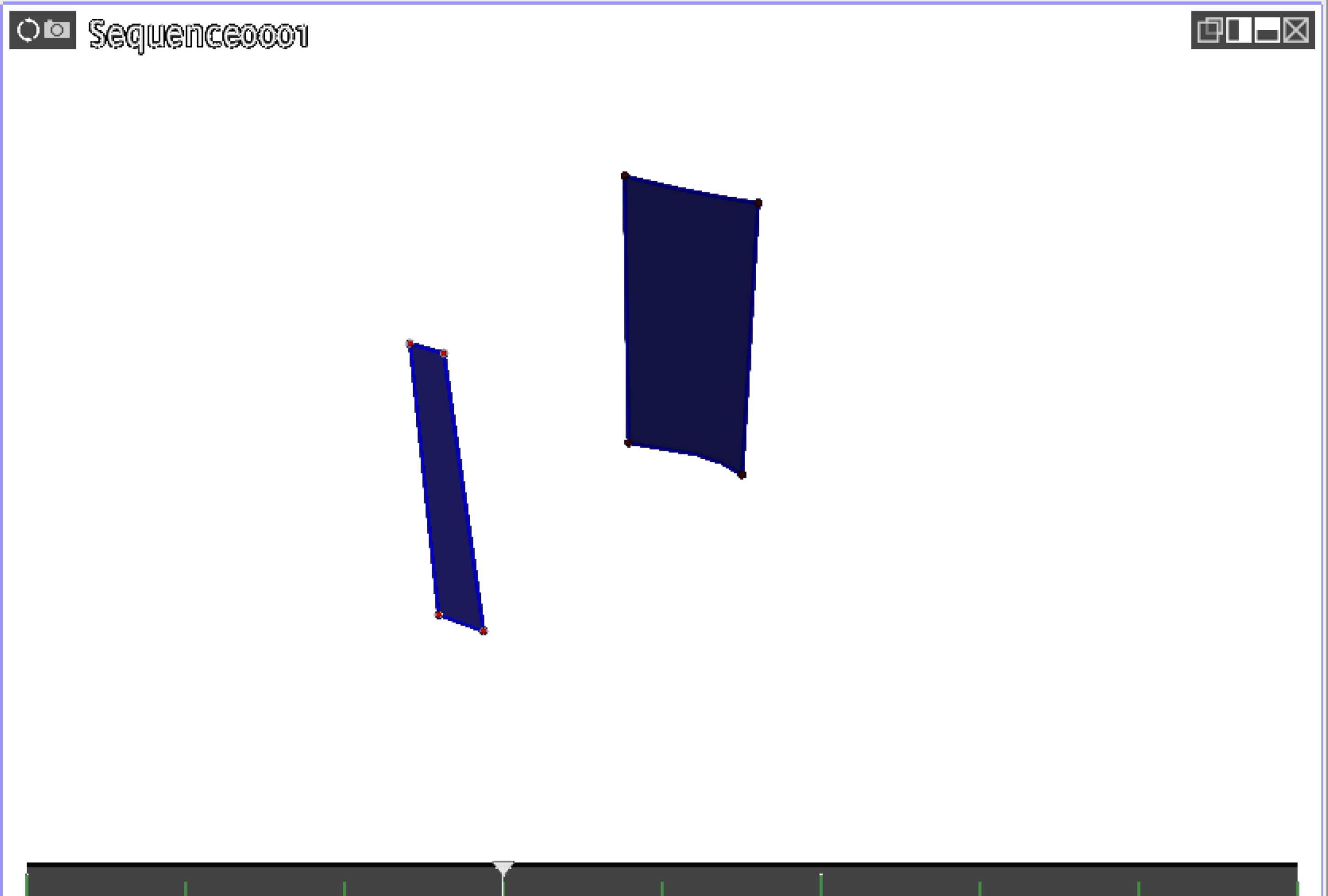} \\
		\includegraphics[width=0.23\textwidth]{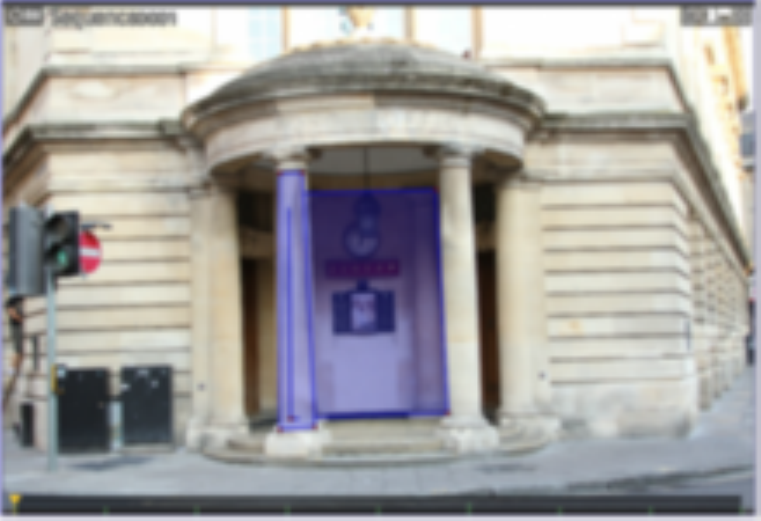} & 
		\includegraphics[width=0.23\textwidth]{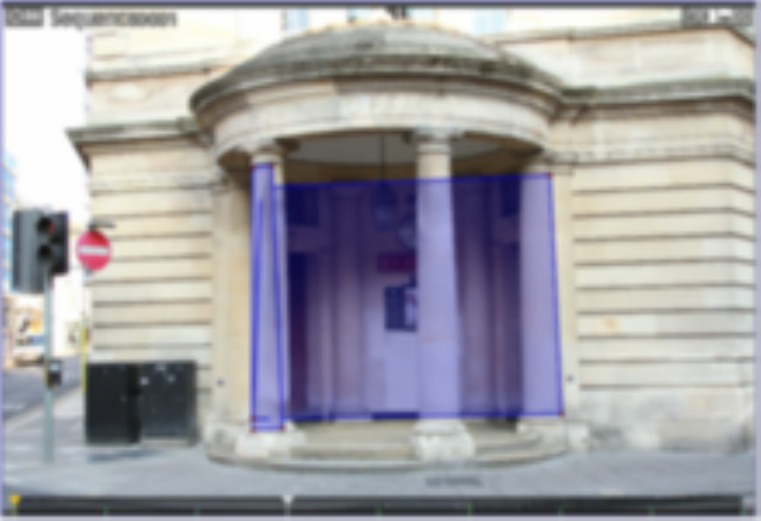} &  
		\includegraphics[width=0.23\textwidth]{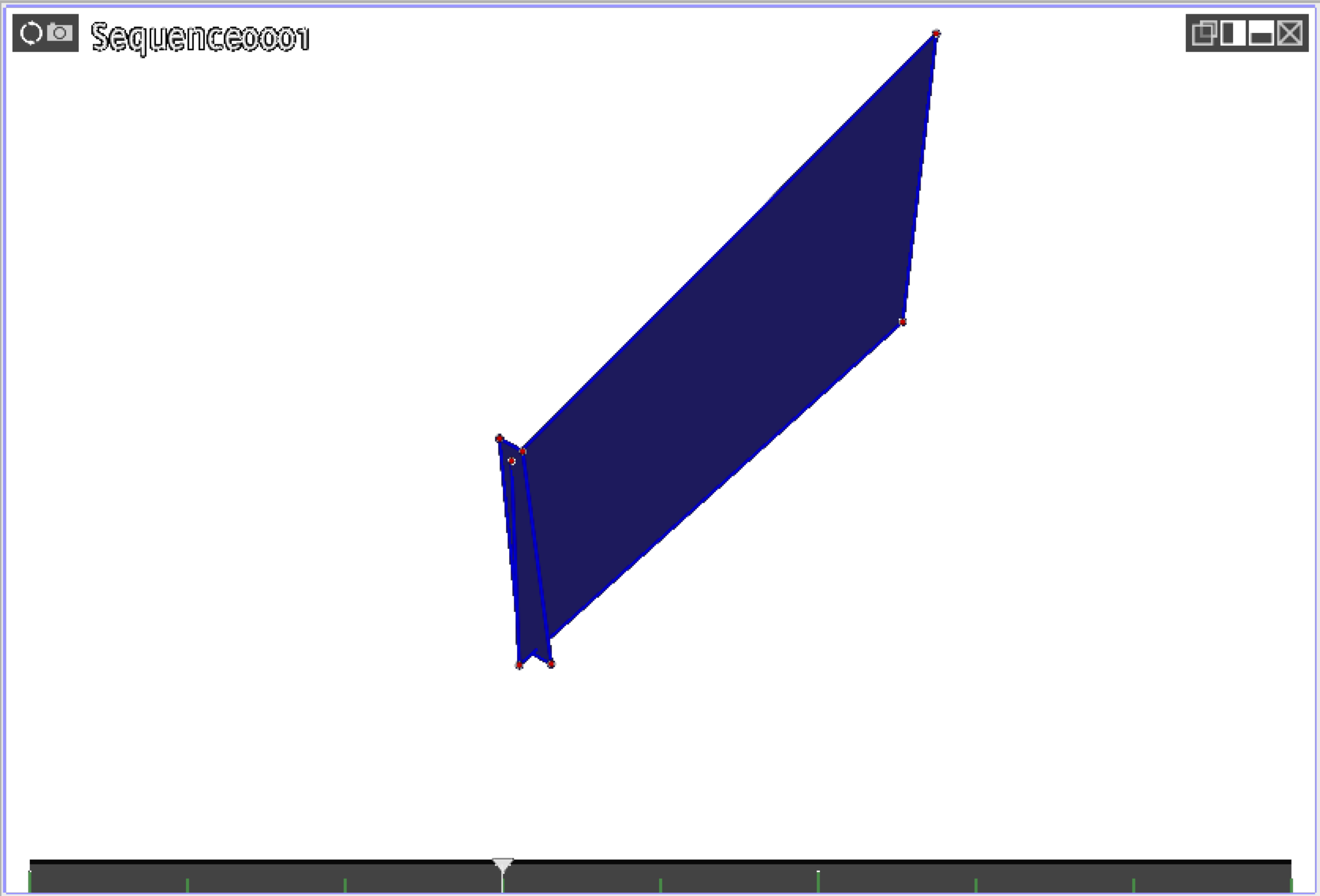}
	\end{tabular}
	\vspace{0.2cm}
	\begin{tabular}{c c c}
		\multicolumn{3}{c}{\textbf{This system}} \\
		View 1 & View 2 & Reconstruction \\
		\includegraphics[width=0.23\textwidth]{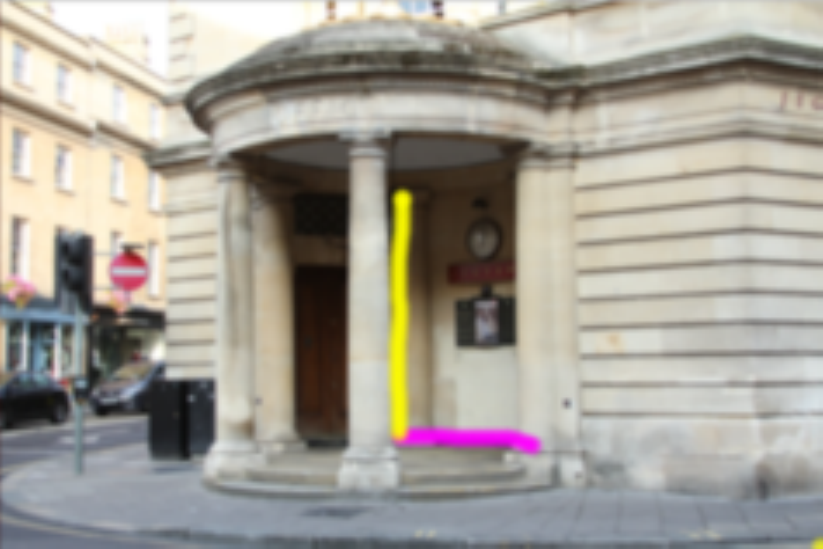} & 
		\includegraphics[width=0.23\textwidth]{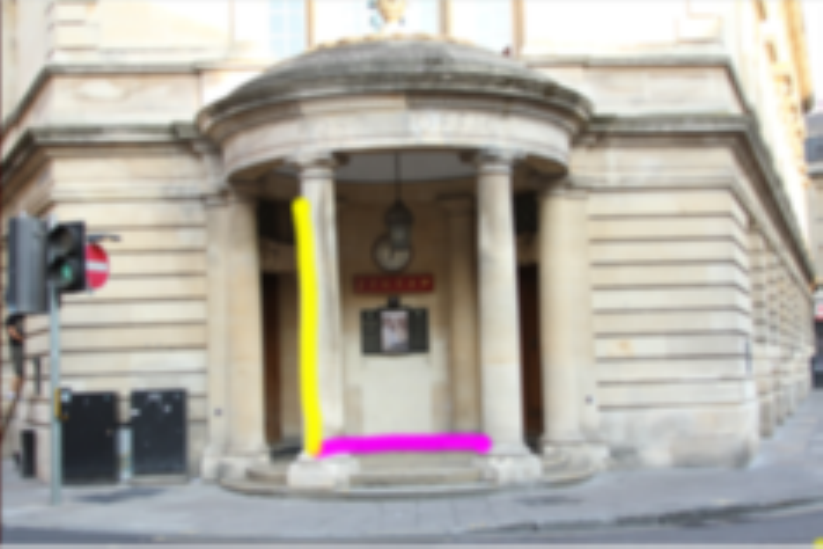} &  
		\includegraphics[width=0.23\textwidth]{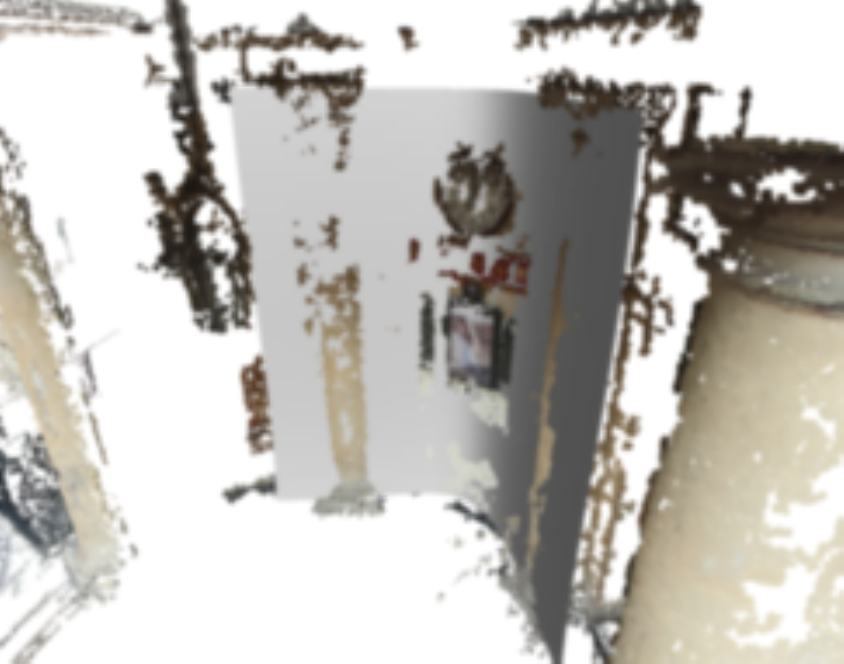} 
	\end{tabular}
	\caption{Example output from video trace. The first row shows a successful surface reconstruction of an occluded face, and the second row shows an unsuccessful reconstruction of the same face. The final row is the output from the system presented in this paper.}
	\label{fig:videotracecomp}
\end{figure*}

One of the principal limitations of this system is the way in which the user identifies the 3D surface. Assuming that the camera geometry and point cloud have been accurately reconstructed, the system takes a polygon from a single view and uses the point cloud to infer where the polygon lies in 3D space. The user can then refine the fit by moving control points from different views. This presents a problem if there is a surface which is occluded, such as the post office canopy in Figure \ref{fig:videotracecomp}. The column has been highlighted for comparison. On the top row the occluded surface was identified by creating a polygon on the left hand image, the computed geometry was correct in the second view, and also in the 3D reconstruction on the right hand side. The second row shows this method to fail when the user wants to extract the larger segment on the occluded face. Again, the polygon was identified in the left hand column but is significantly erroneous in both the second view and the 3D reconstruction on the right hand side. The method presented in this paper correctly identifies the surface, even though we have sketched over an occlusion, and is displayed on the bottom row. 

\subsection{Qualitative results}
We provide some final comparisons on a dataset of objects, all of which are man-made objects in an urban environment. The dataset contains building which have a surface structure which is not restricted to planes alone, making a large number of traditional methods inapplicable. 

\begin{figure*}[t!]
	\centering
	\def\arraystretch{3}
	\begin{tabular}{>{\centering\arraybackslash}p{0.01\linewidth}  c c c c}
		& Point cloud & Poisson reconstruction & Primitives & Closure \\
		\rotatebox{90}{\hspace{0.3in}{The cross} } &
		\includegraphics[width=0.2\linewidth]{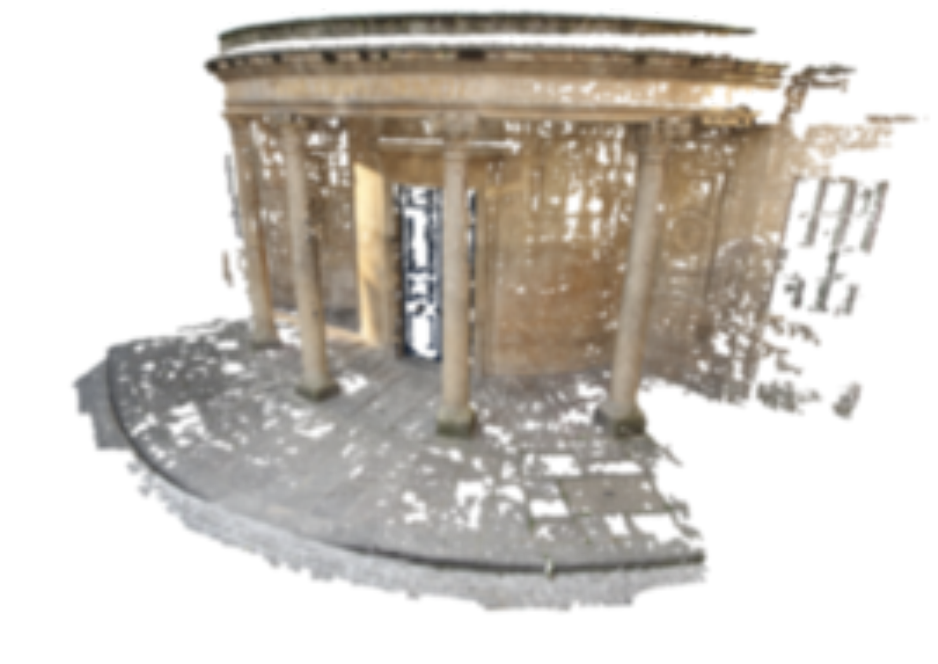} &
		 \includegraphics[width=0.2\linewidth]{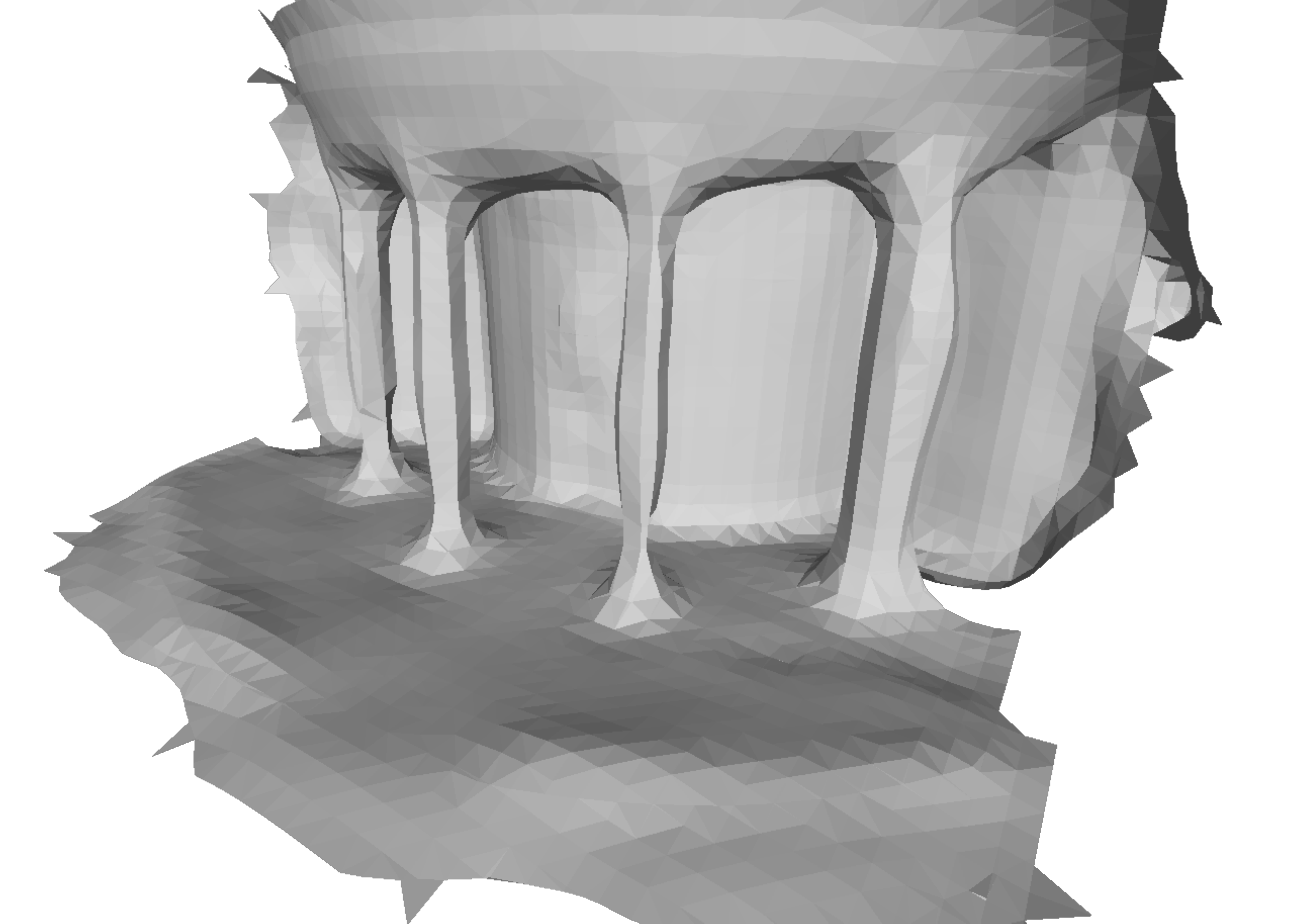} &
		 \includegraphics[width=0.2\linewidth]{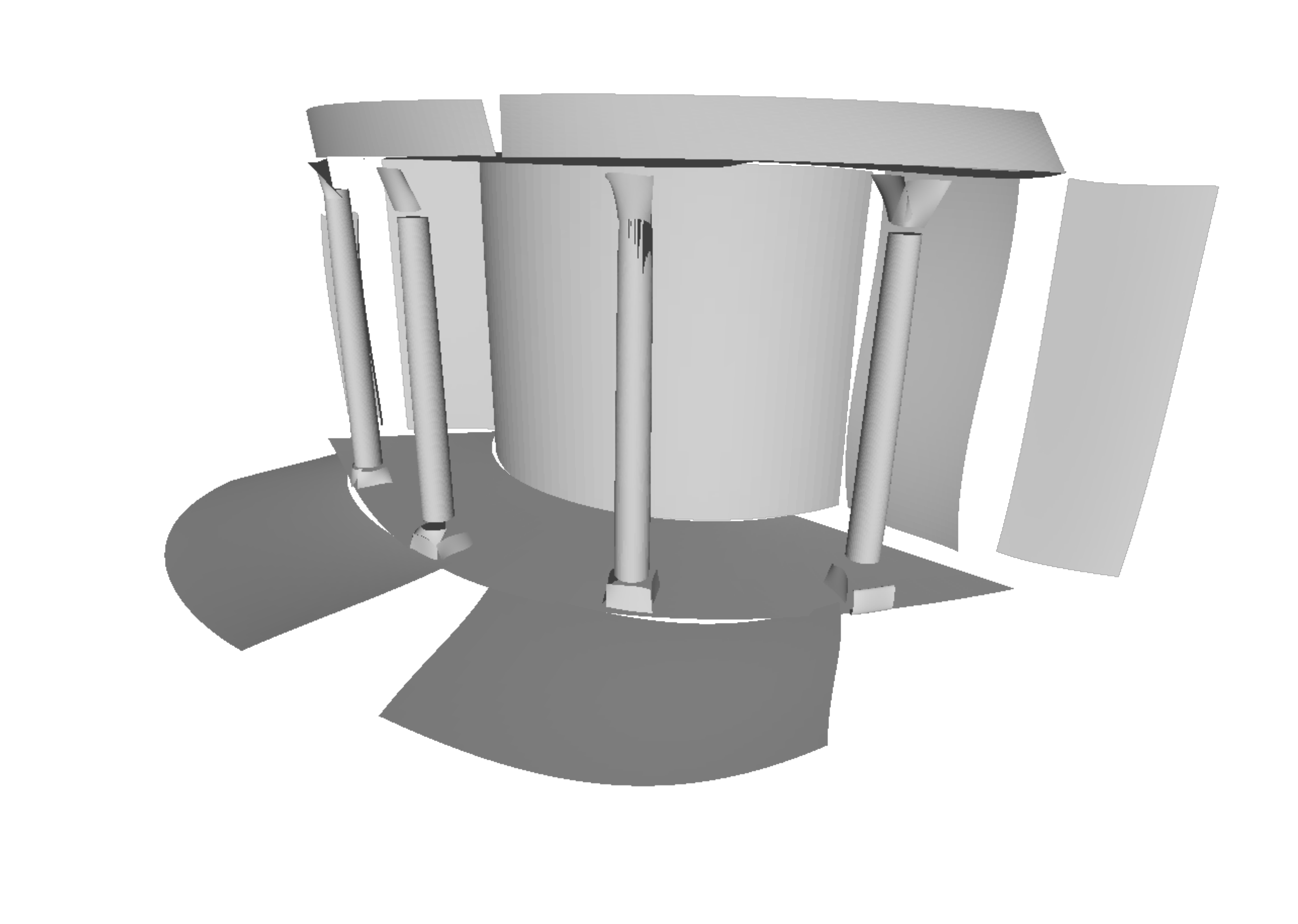} &
		 \includegraphics[width=0.2\linewidth]{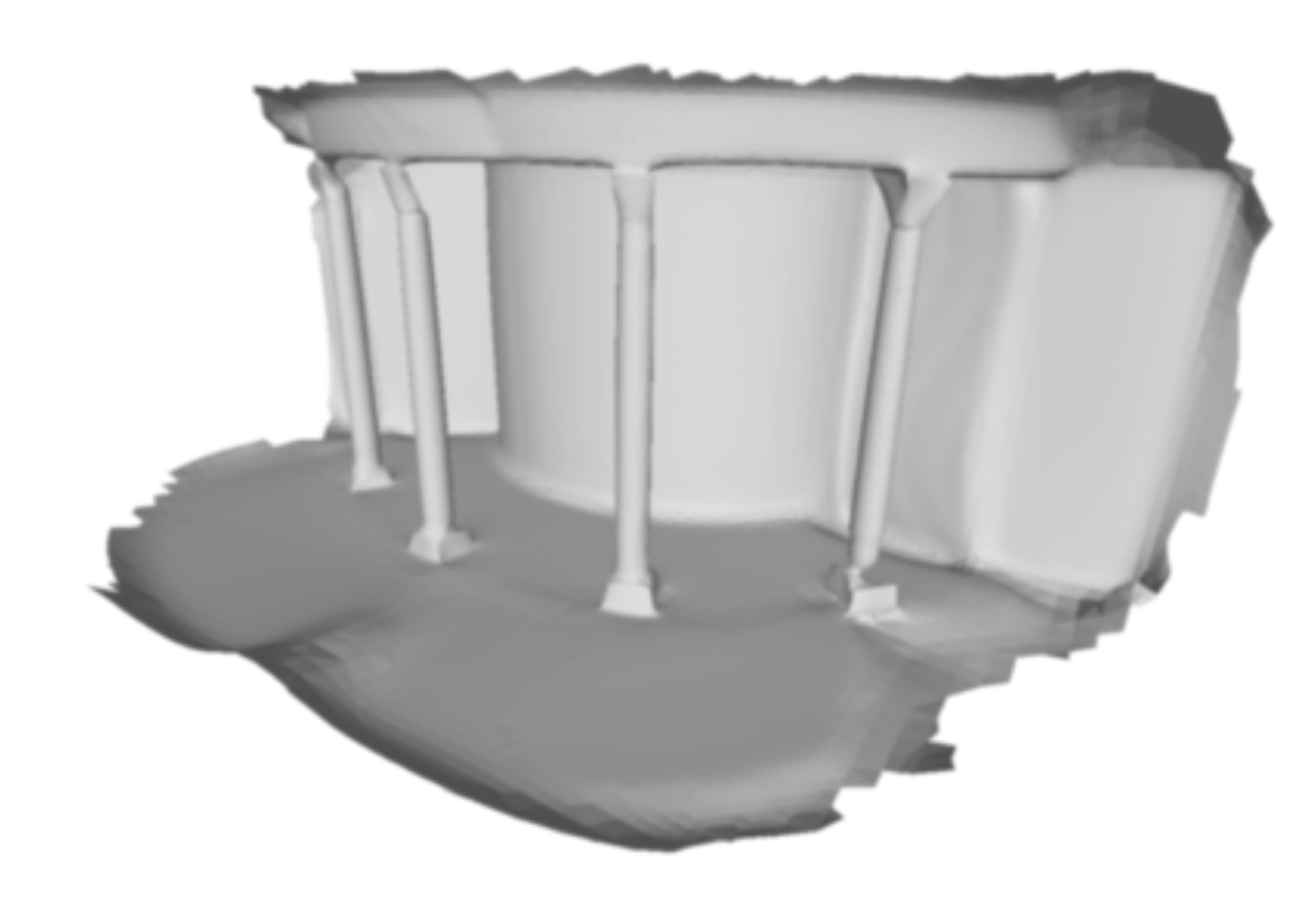} \\
		\rotatebox{90}{\hspace{0in}{The post office} } &
		 \includegraphics[width=0.2\linewidth]{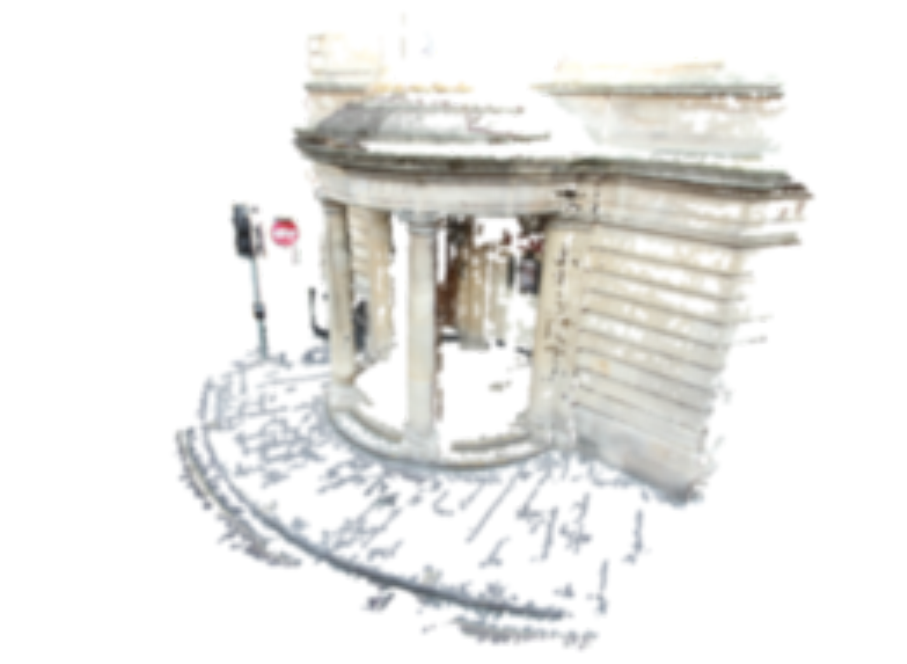} &
		 \includegraphics[width=0.2\linewidth]{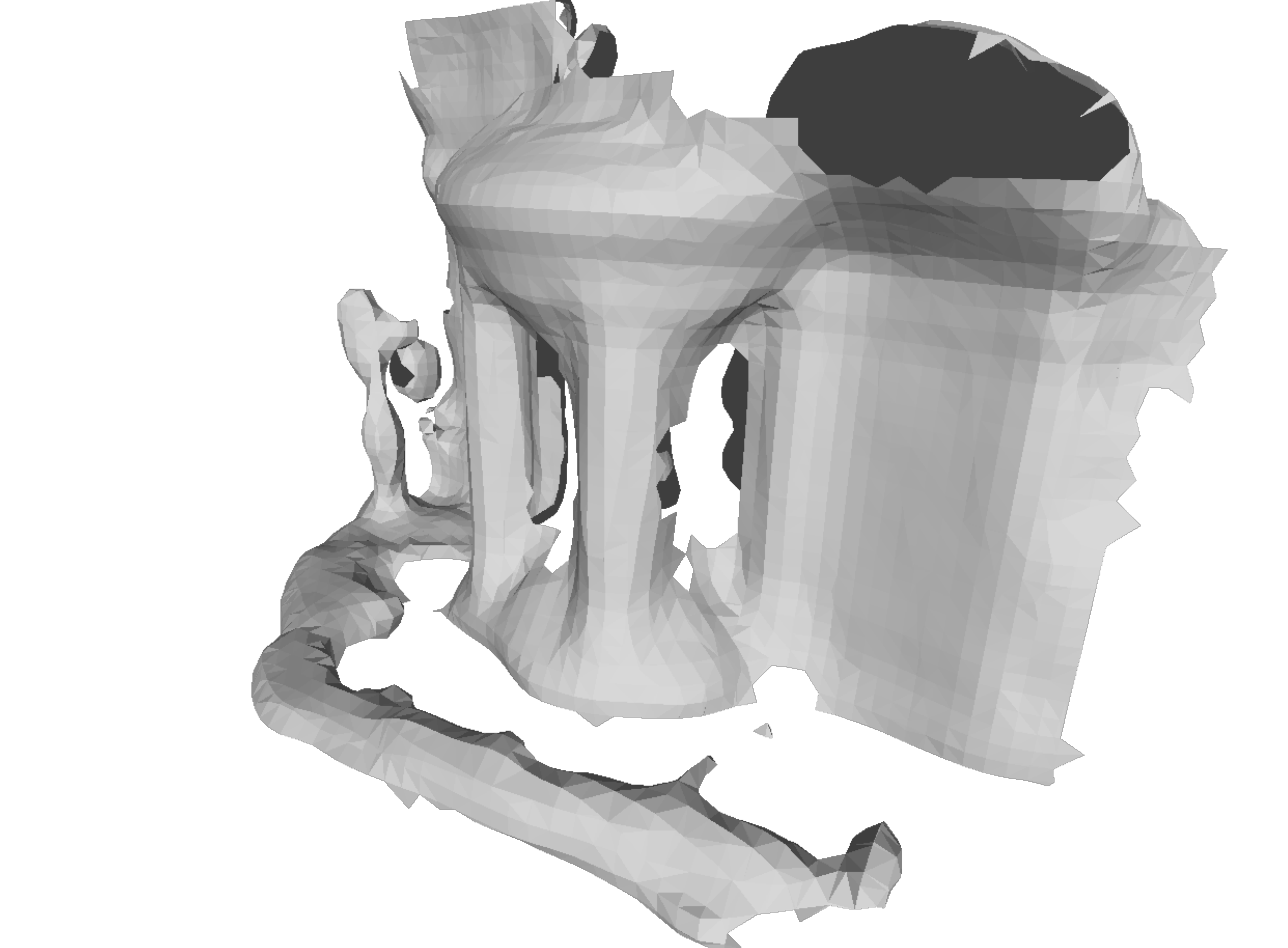} &
		 \includegraphics[width=0.2\linewidth]{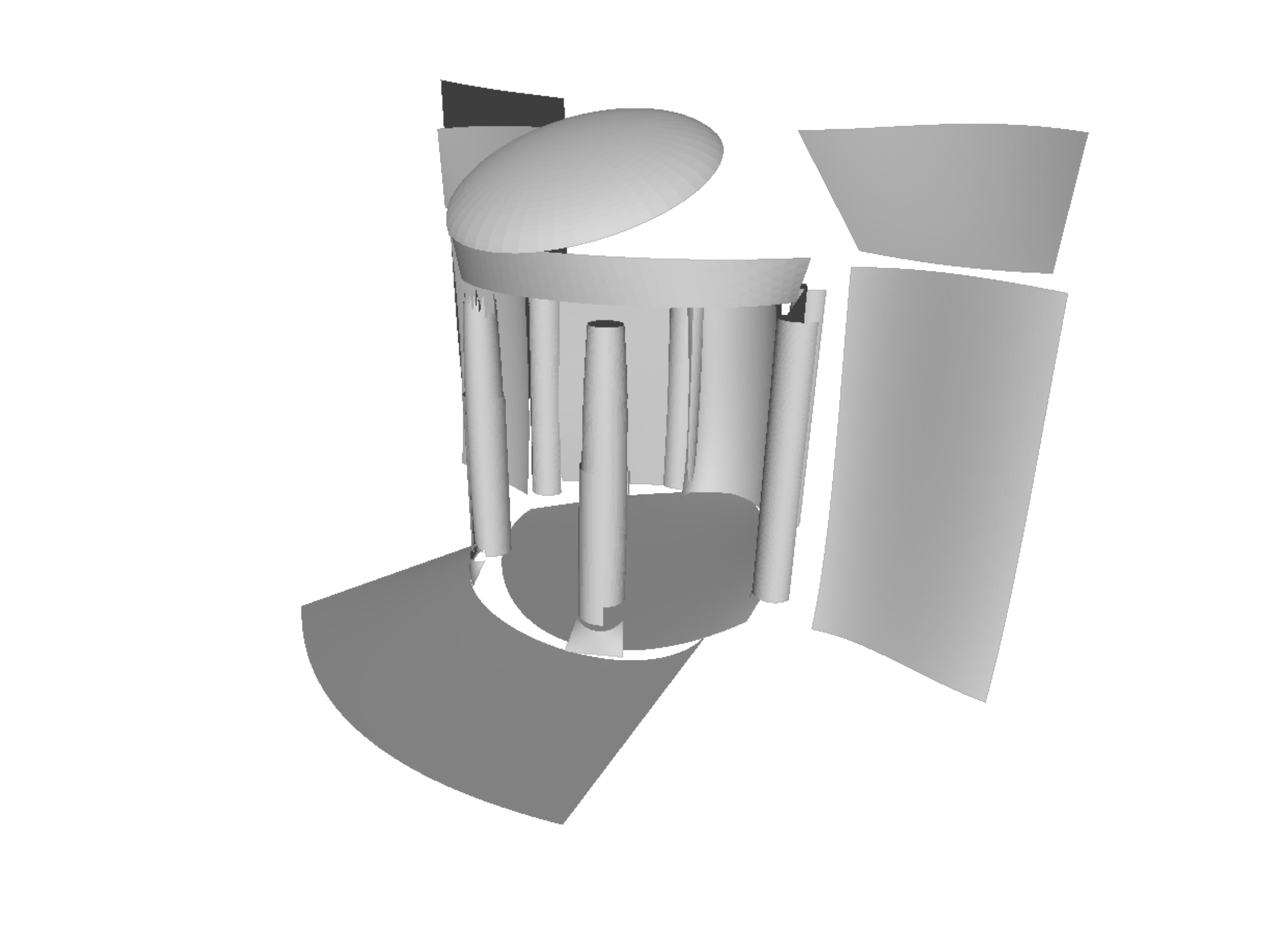} &
		 \includegraphics[width=0.2\linewidth]{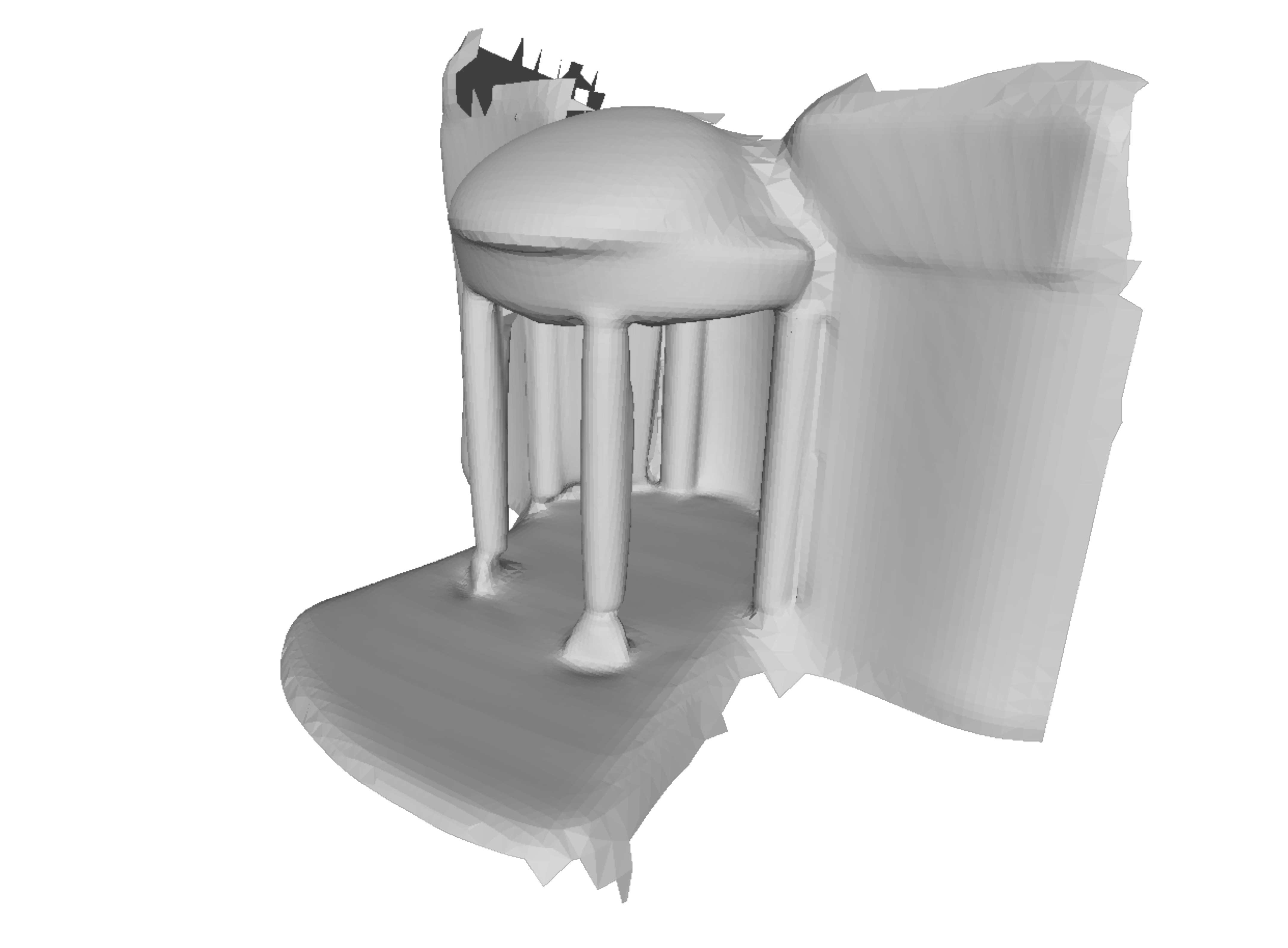} \\
		\rotatebox{90}{\hspace{0in}{Mantova Basilica} } &
		 \includegraphics[width=0.2\linewidth]{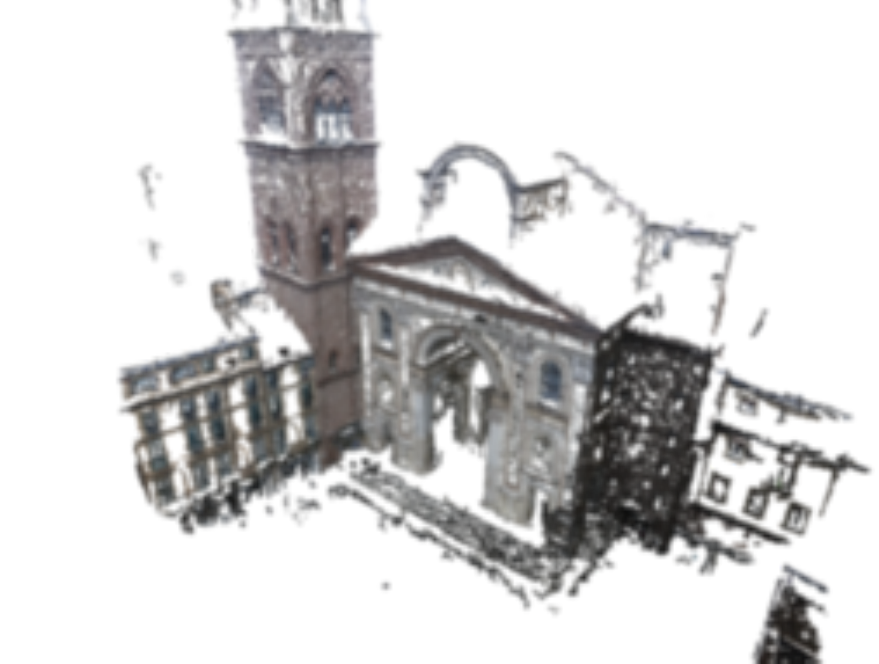} &
		 \includegraphics[width=0.2\linewidth]{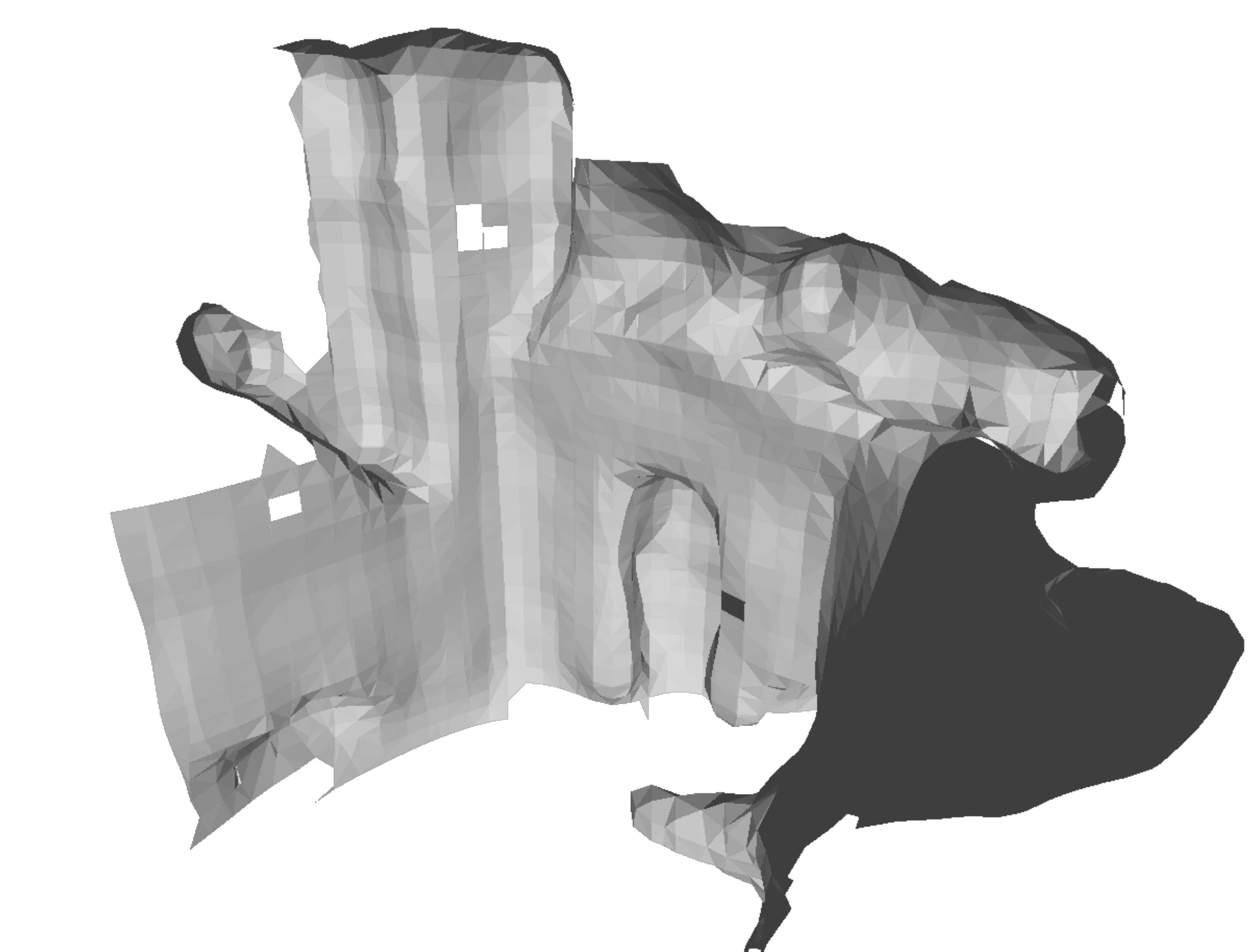} &
		 \includegraphics[width=0.2\linewidth]{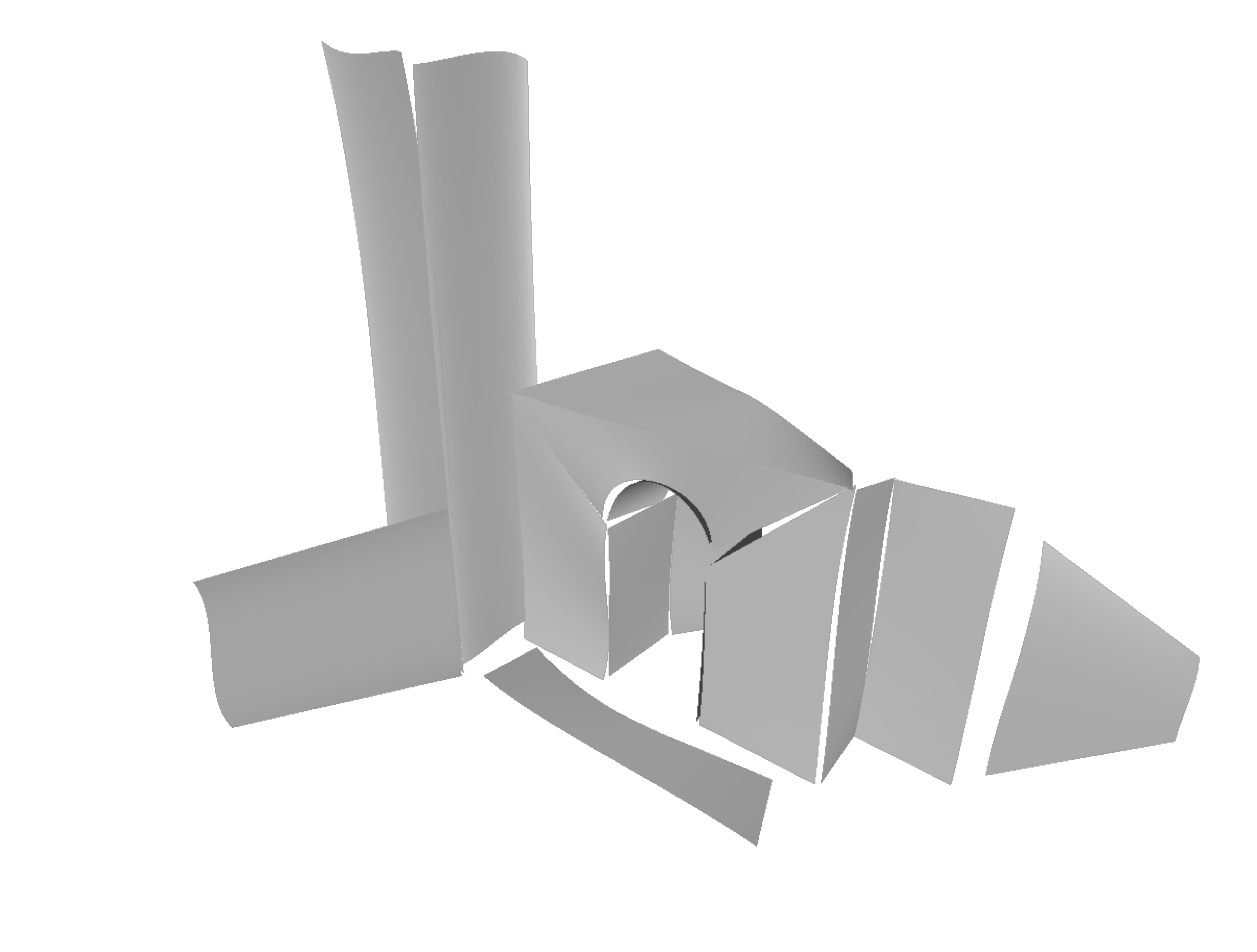} &
		 \includegraphics[width=0.2\linewidth]{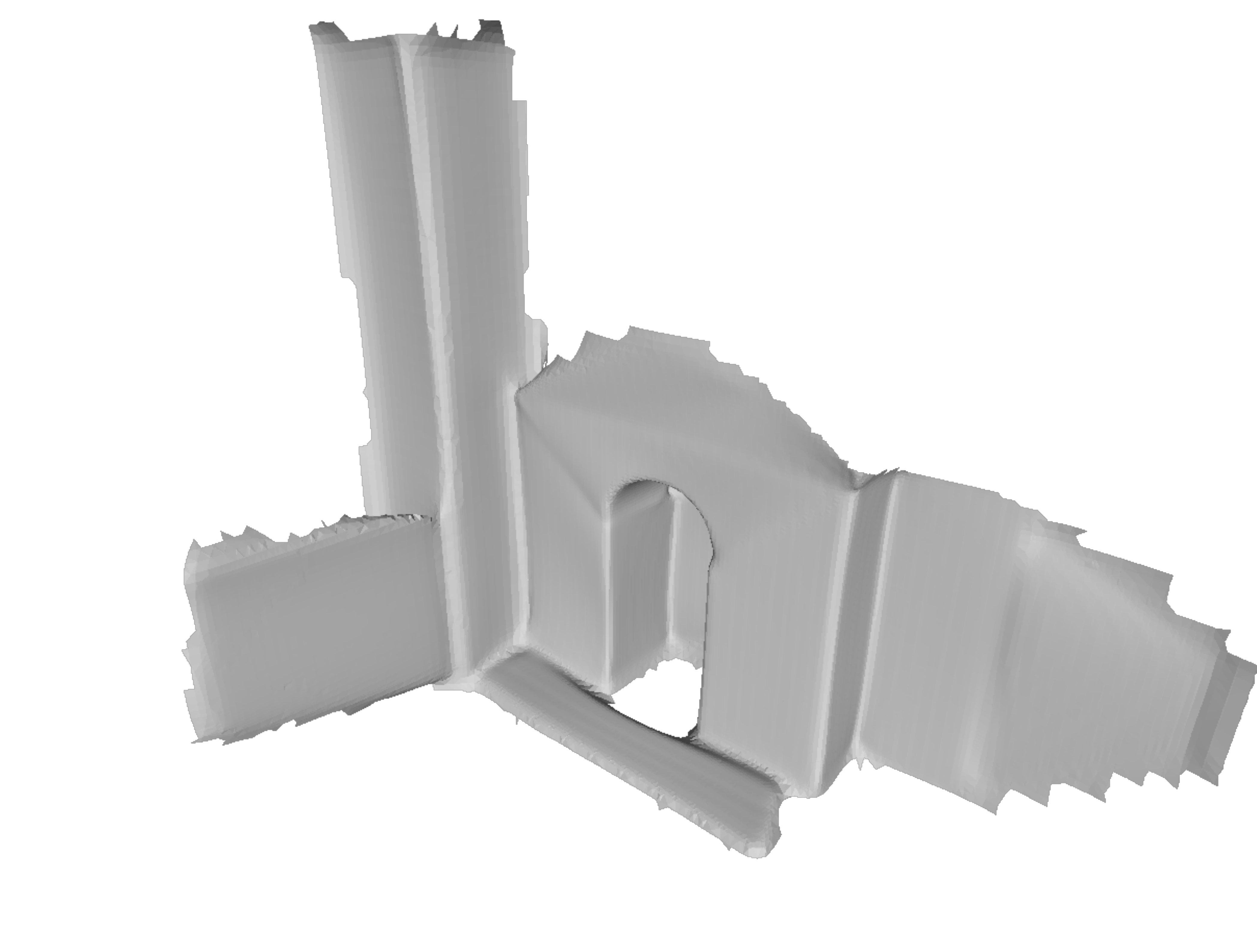}  \\
		\rotatebox{90}{\hspace{0.1in}{Phone booth} } &
		 \includegraphics[width=0.2\linewidth]{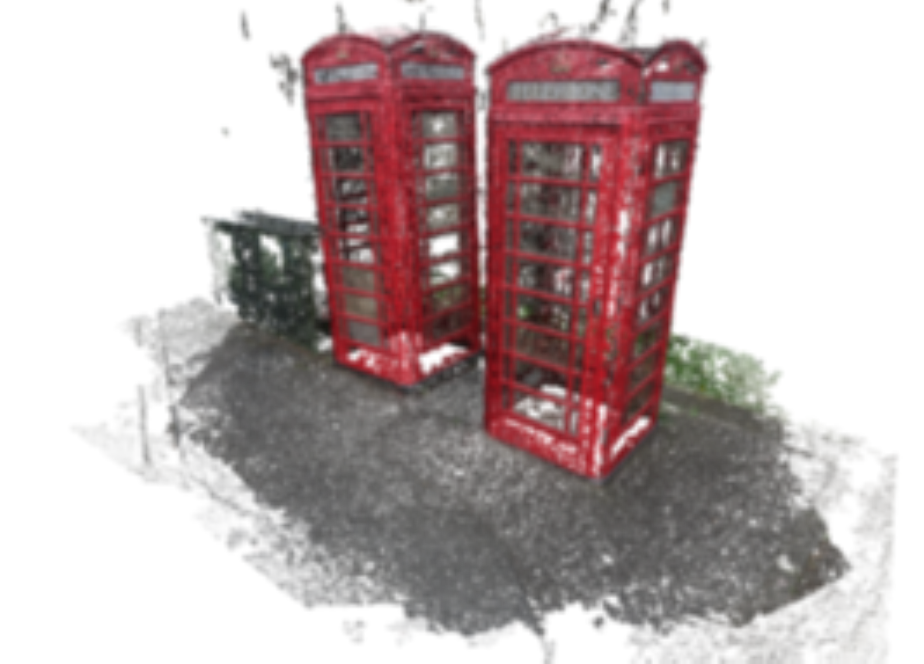} &
		 \includegraphics[width=0.2\linewidth]{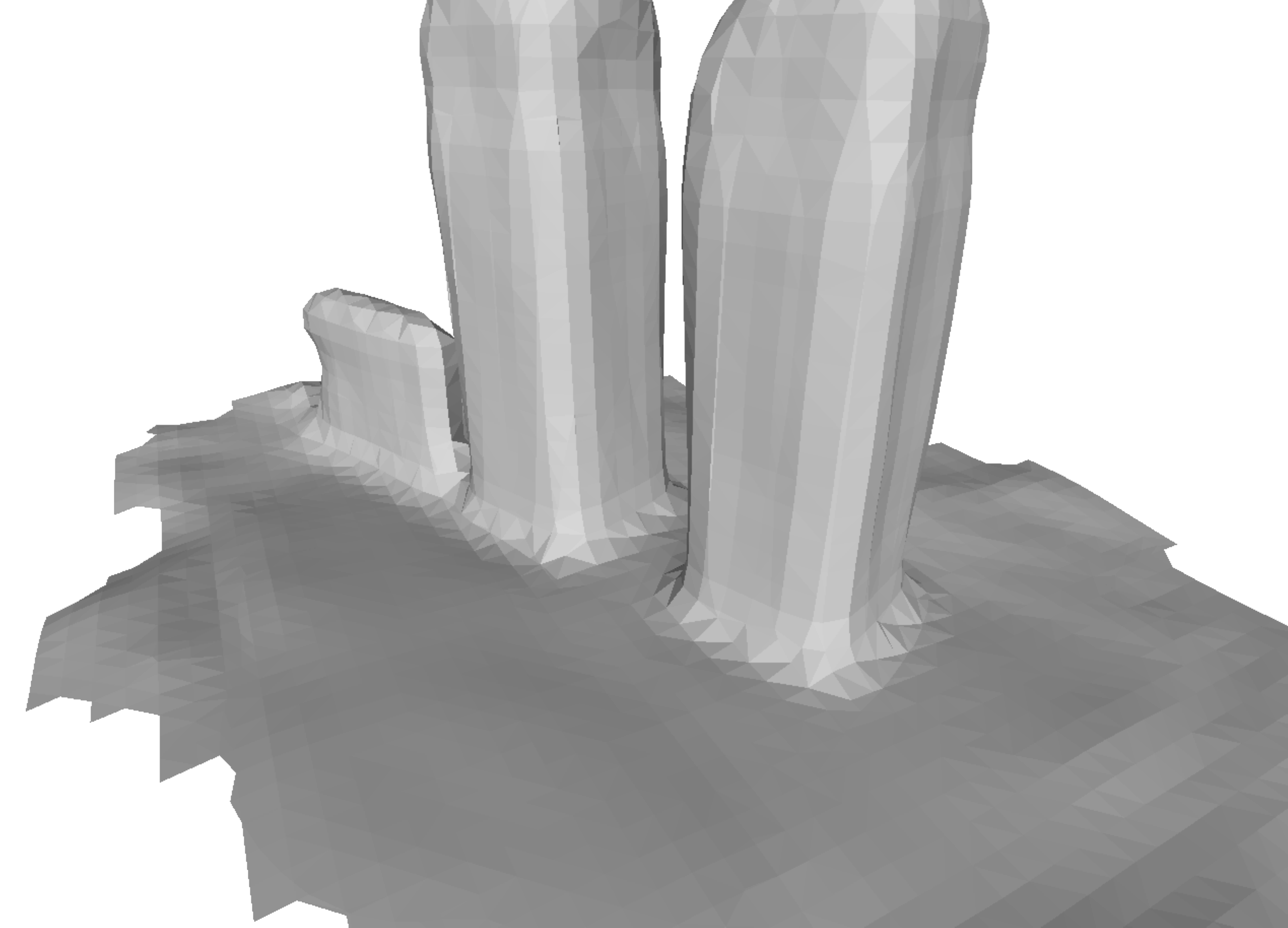} &
		 \includegraphics[width=0.2\linewidth]{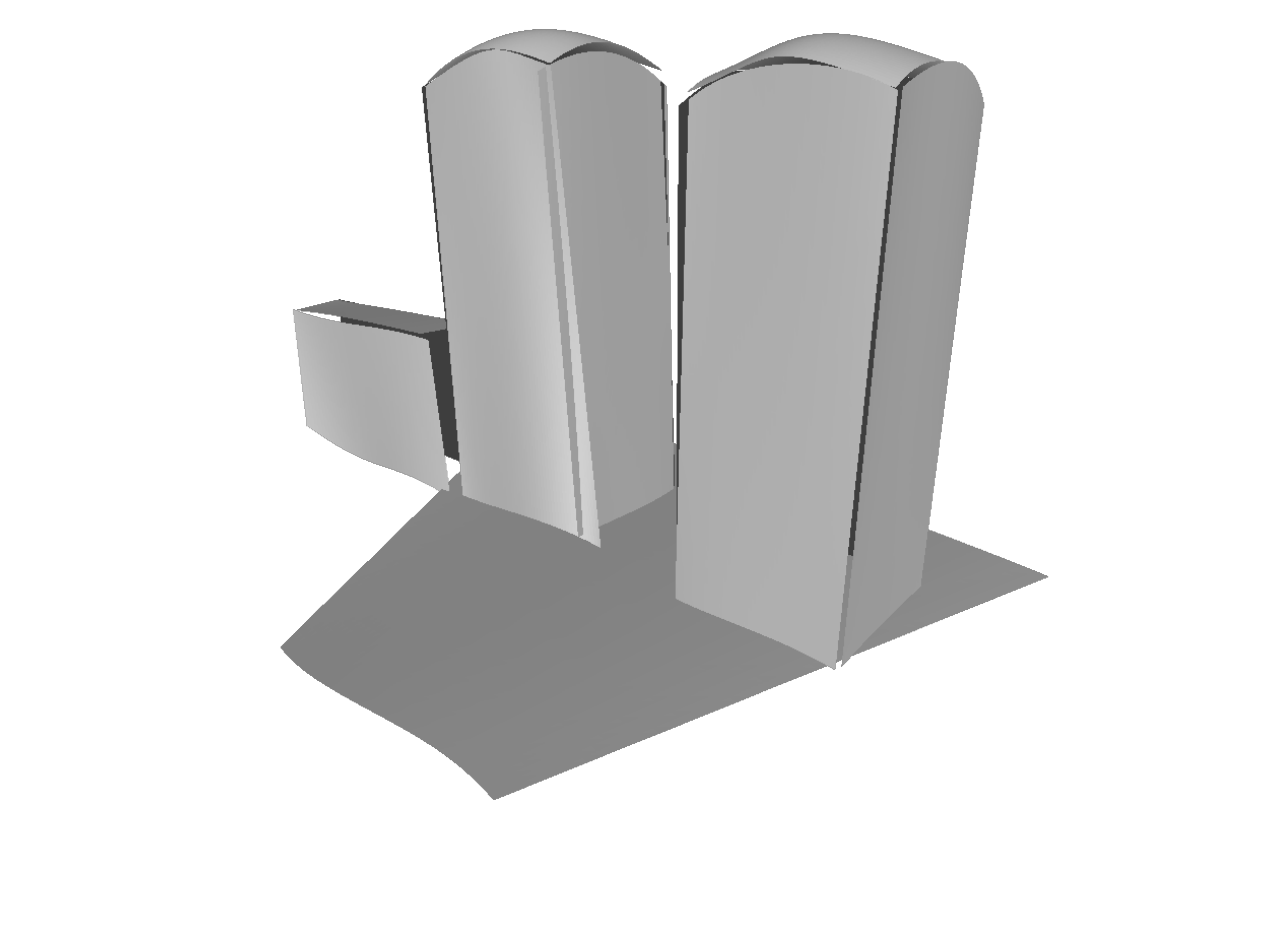} &
		 \includegraphics[width=0.2\linewidth]{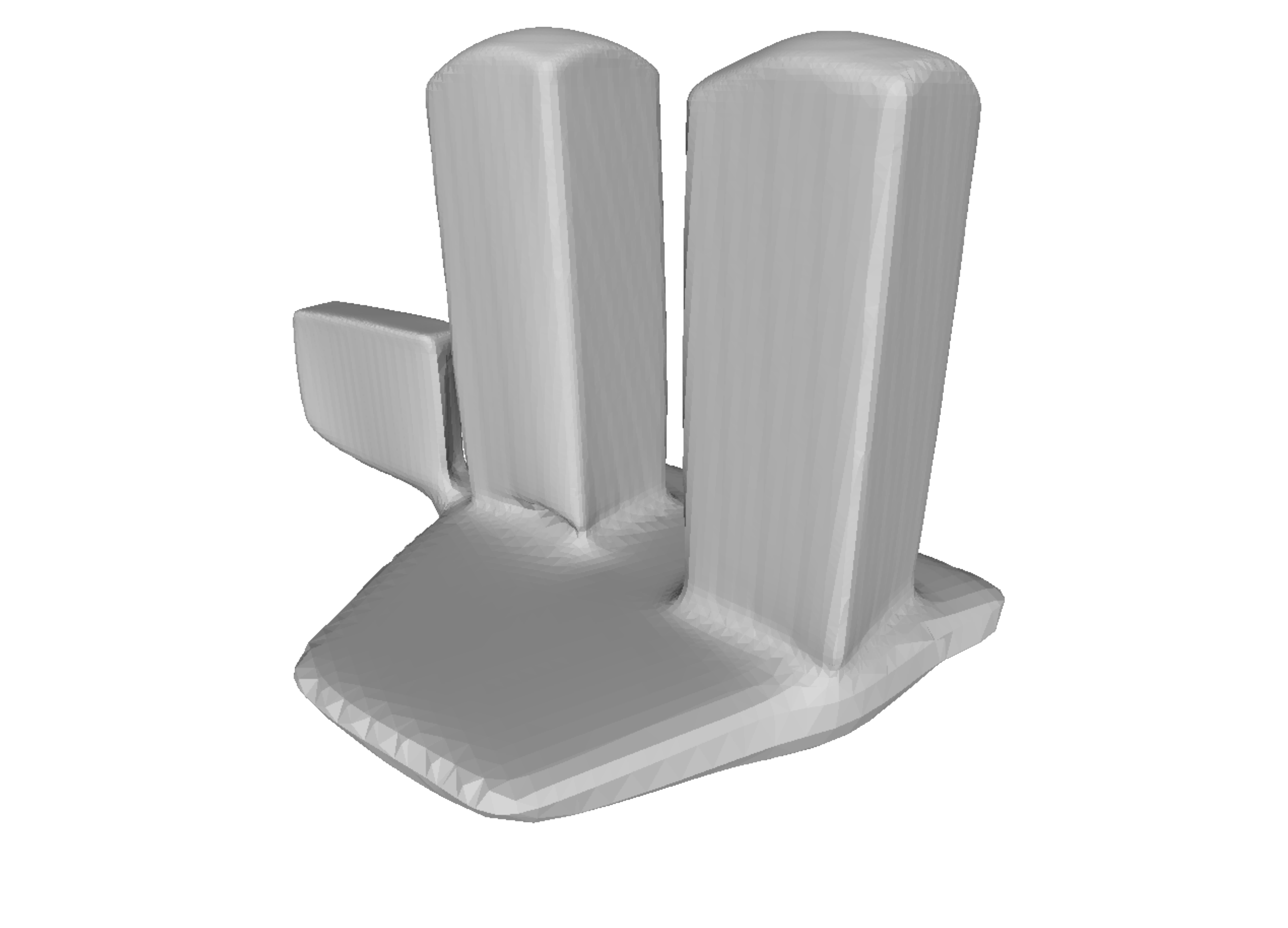} \\
		\rotatebox{90}{\hspace{0in}{Mantova Piazza} } &
		 \includegraphics[width=0.2\linewidth]{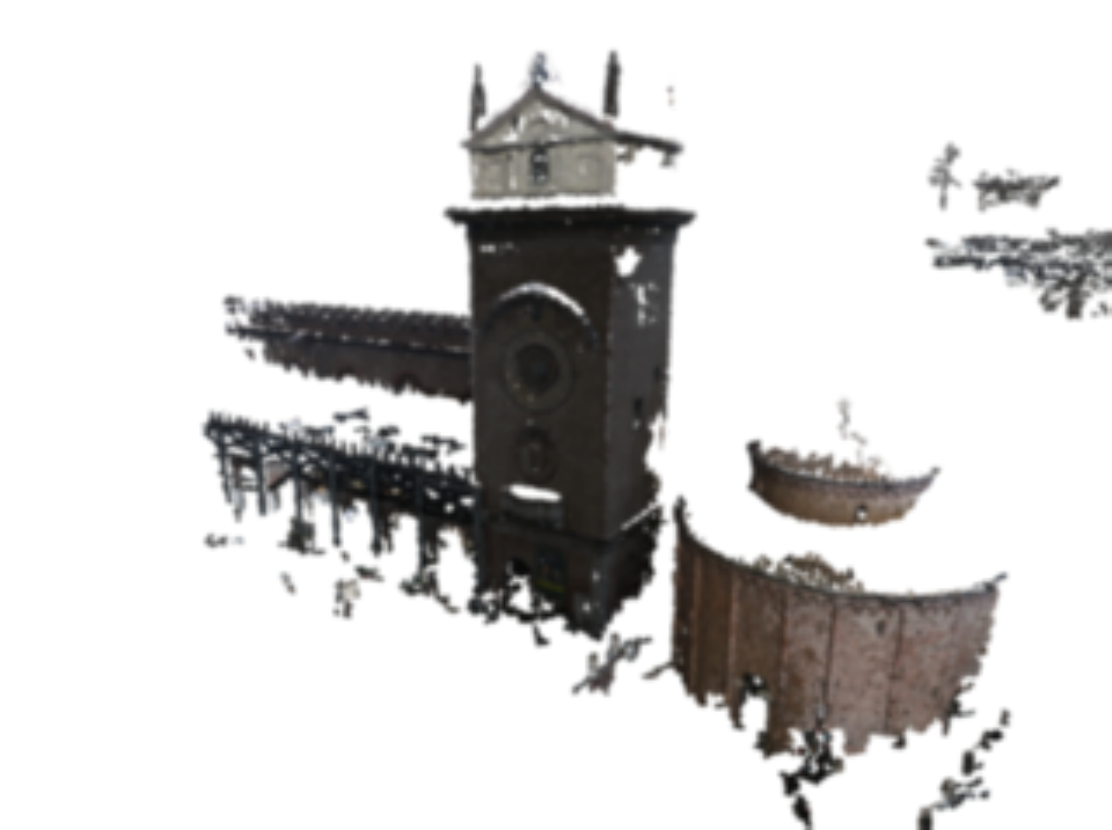} &
		 \includegraphics[width=0.2\linewidth]{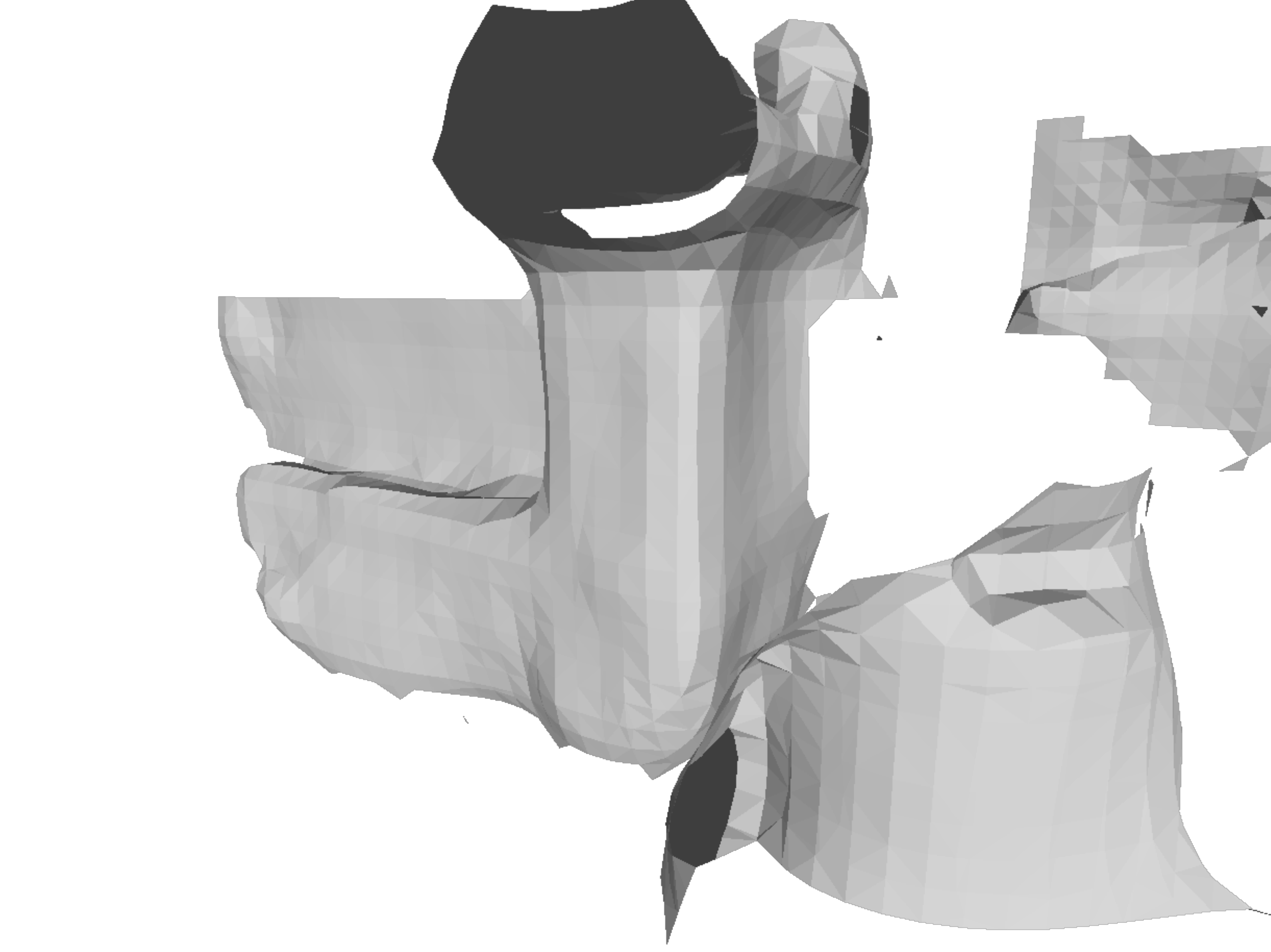} &
		 \includegraphics[width=0.2\linewidth]{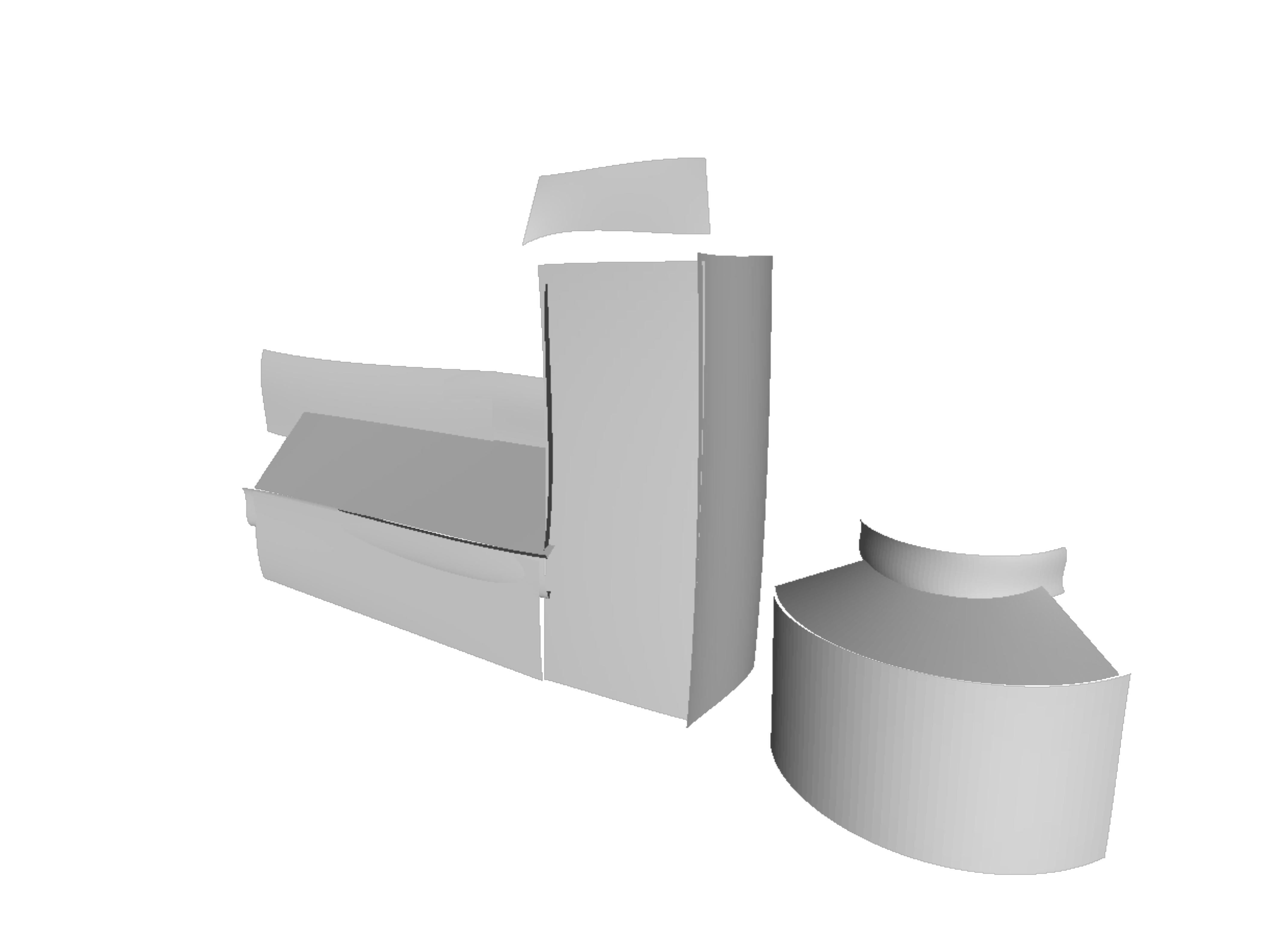} &
		 \includegraphics[width=0.2\linewidth]{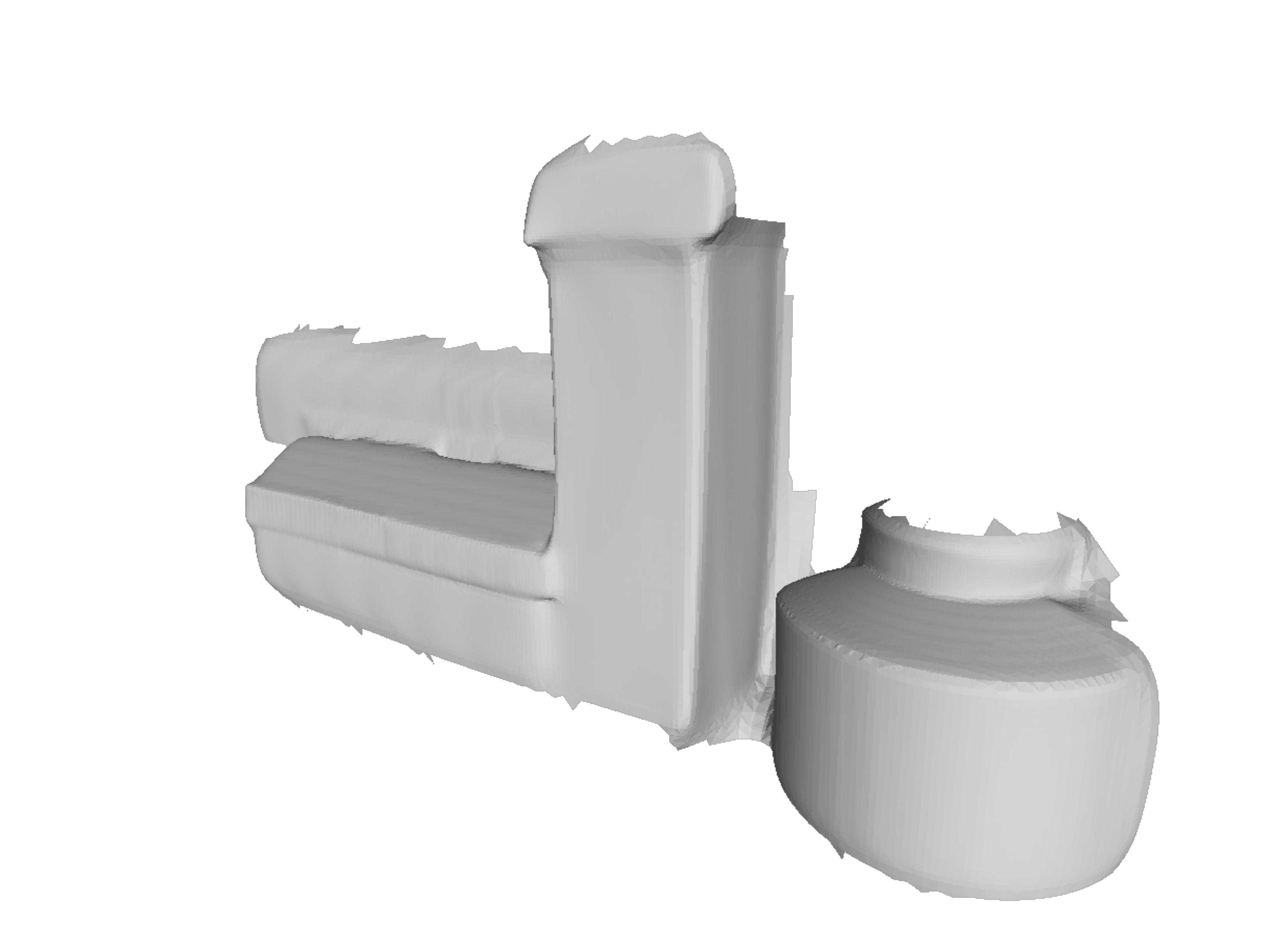} \\
		\rotatebox{90}{\hspace{0.3in}{Post box} } &
		 \includegraphics[width=0.2\linewidth]{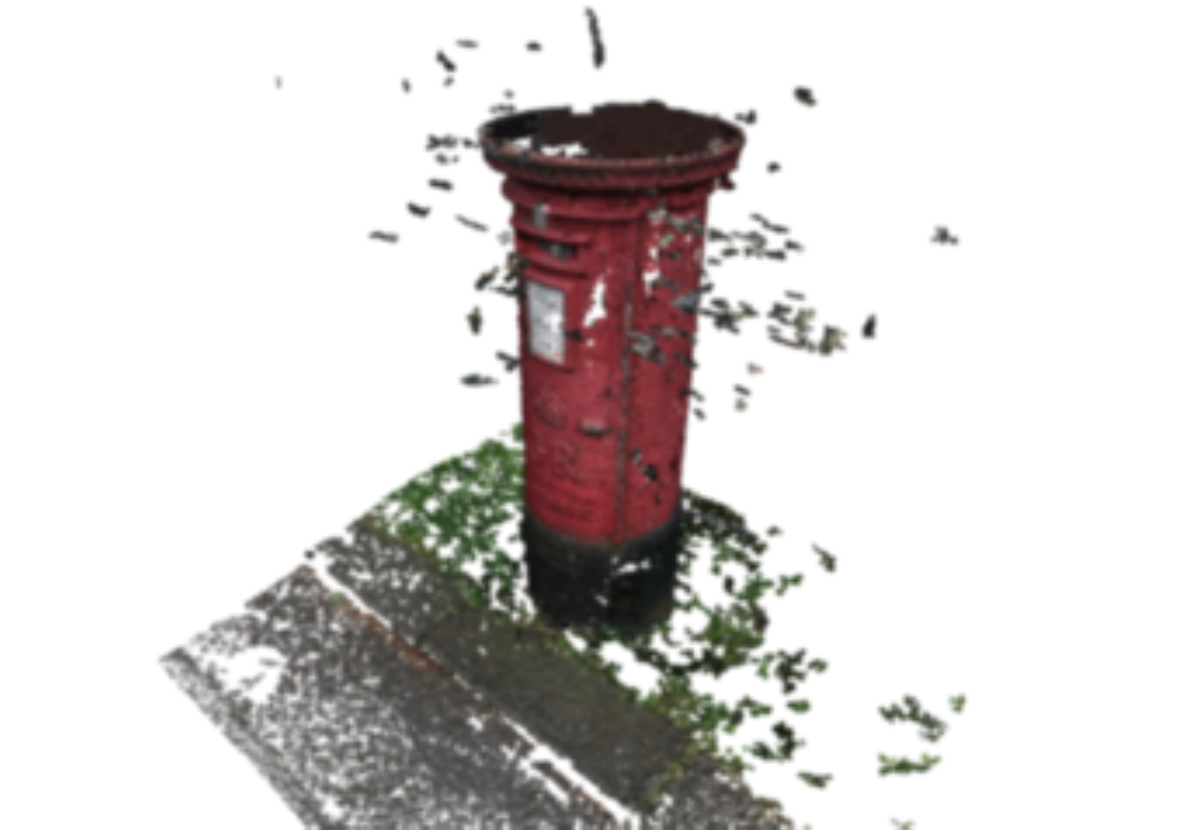} &
		 \includegraphics[width=0.2\linewidth]{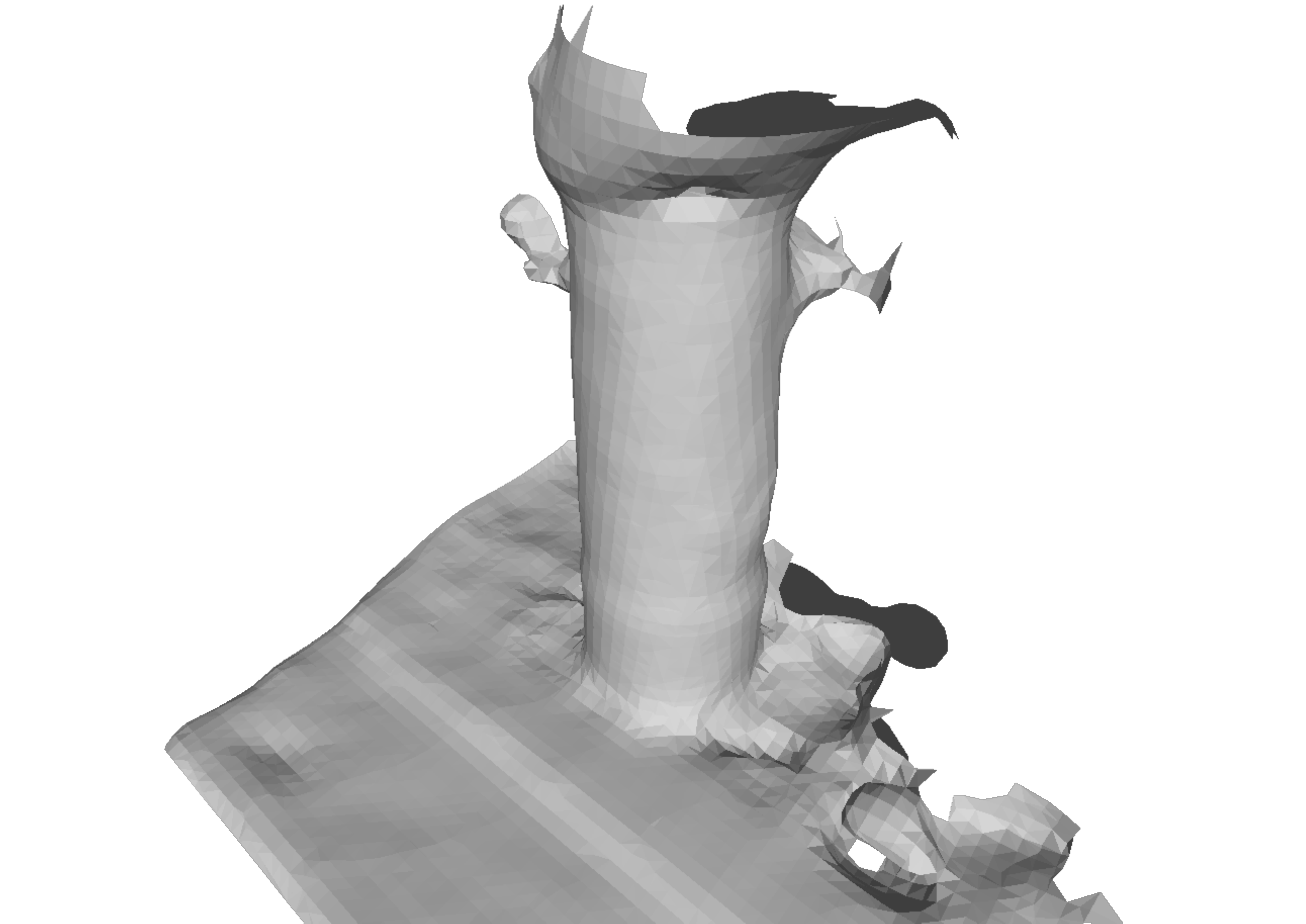} &
		 \includegraphics[width=0.2\linewidth]{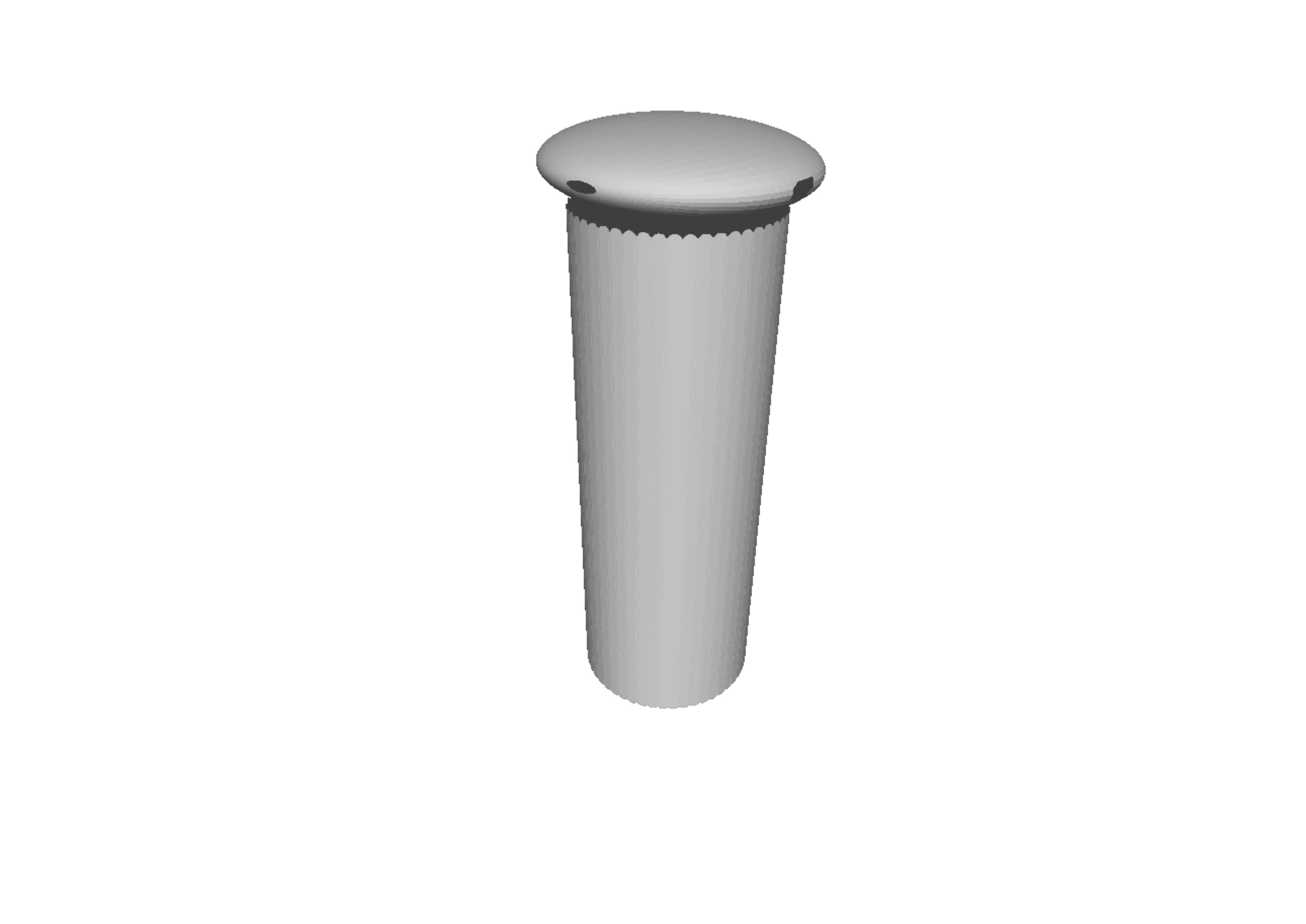} &
		 \includegraphics[width=0.2\linewidth]{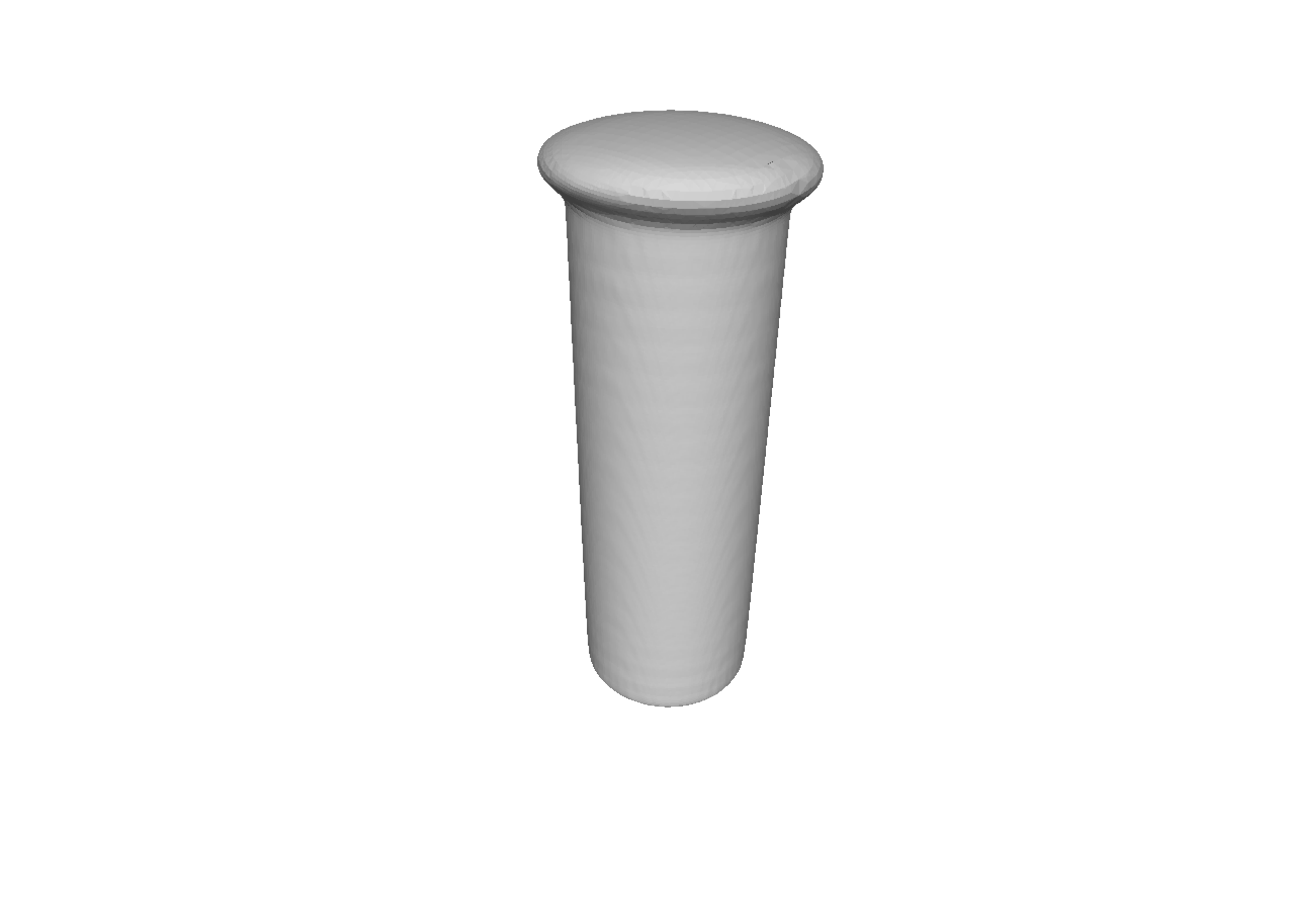}
	\end{tabular}
	\caption{Some results from the method presented in this paper after fitting the primitives and applying surface closure. The far left shows the original point cloud, the second column shows the results of normal estimation and Poisson reconstruction, and the right hand side shows the extracted primitives, and the result of applying a Poisson reconstruction to the output. }
	\label{fig:finalresults}
\end{figure*}

\begin{figure*}[t!]
	\centering
	\def\arraystretch{3}
	\begin{tabular}{c c c}
		\includegraphics[width=0.33\linewidth]{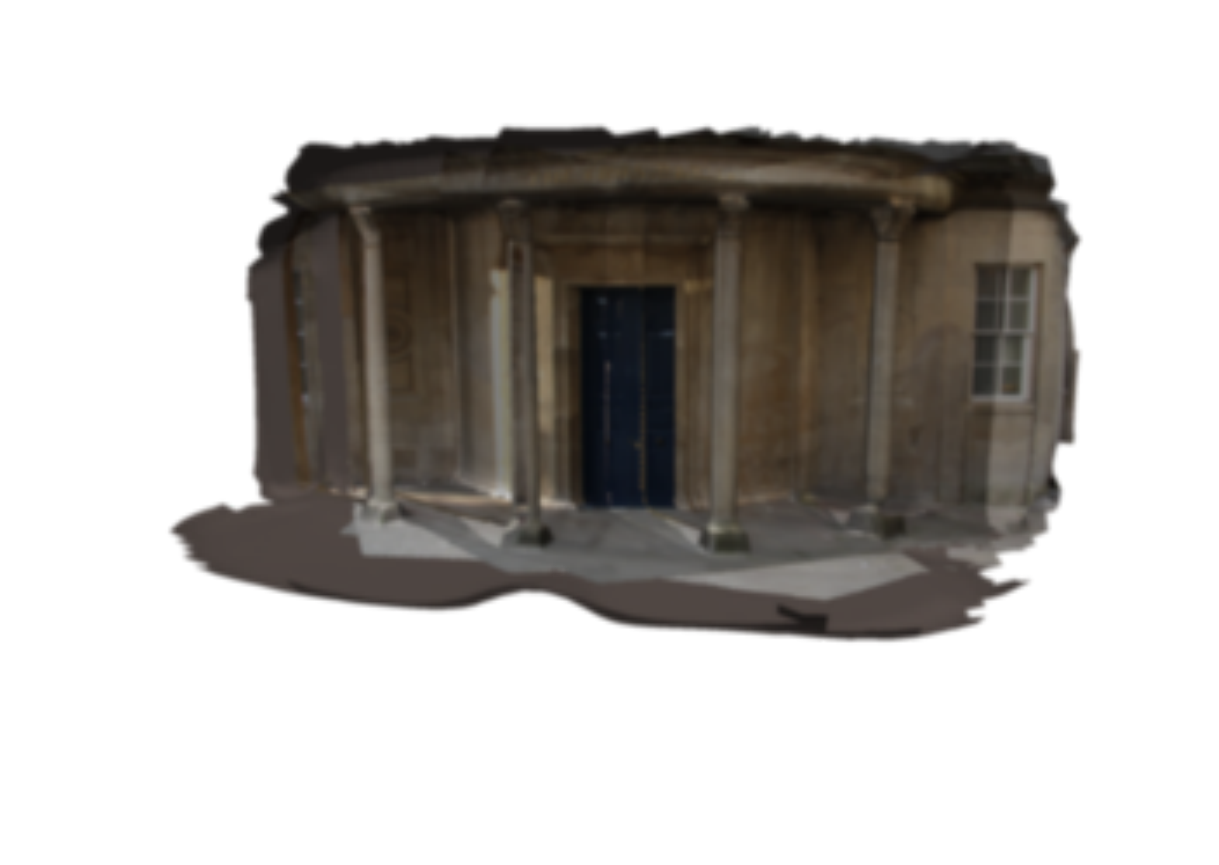} &
		 \includegraphics[width=0.33\linewidth]{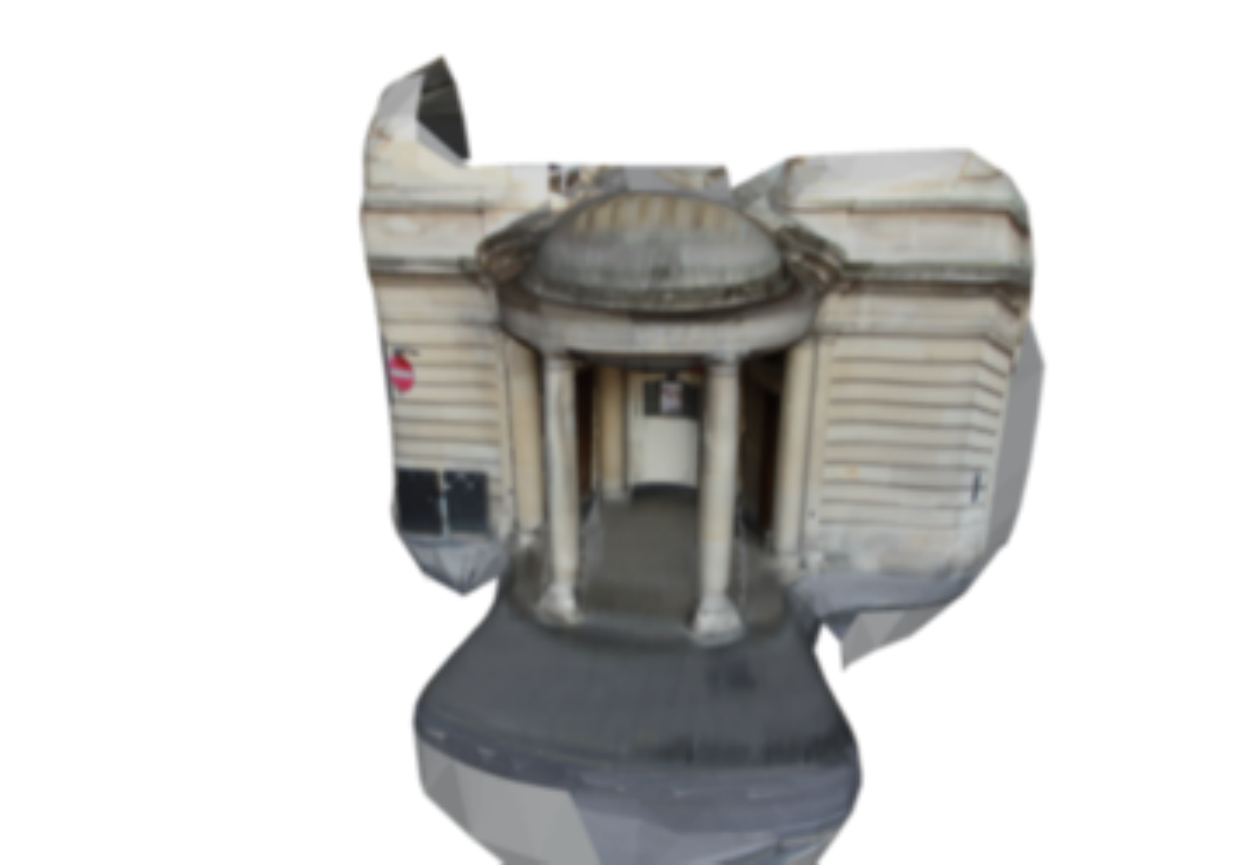} &
		 \includegraphics[width=0.33\linewidth]{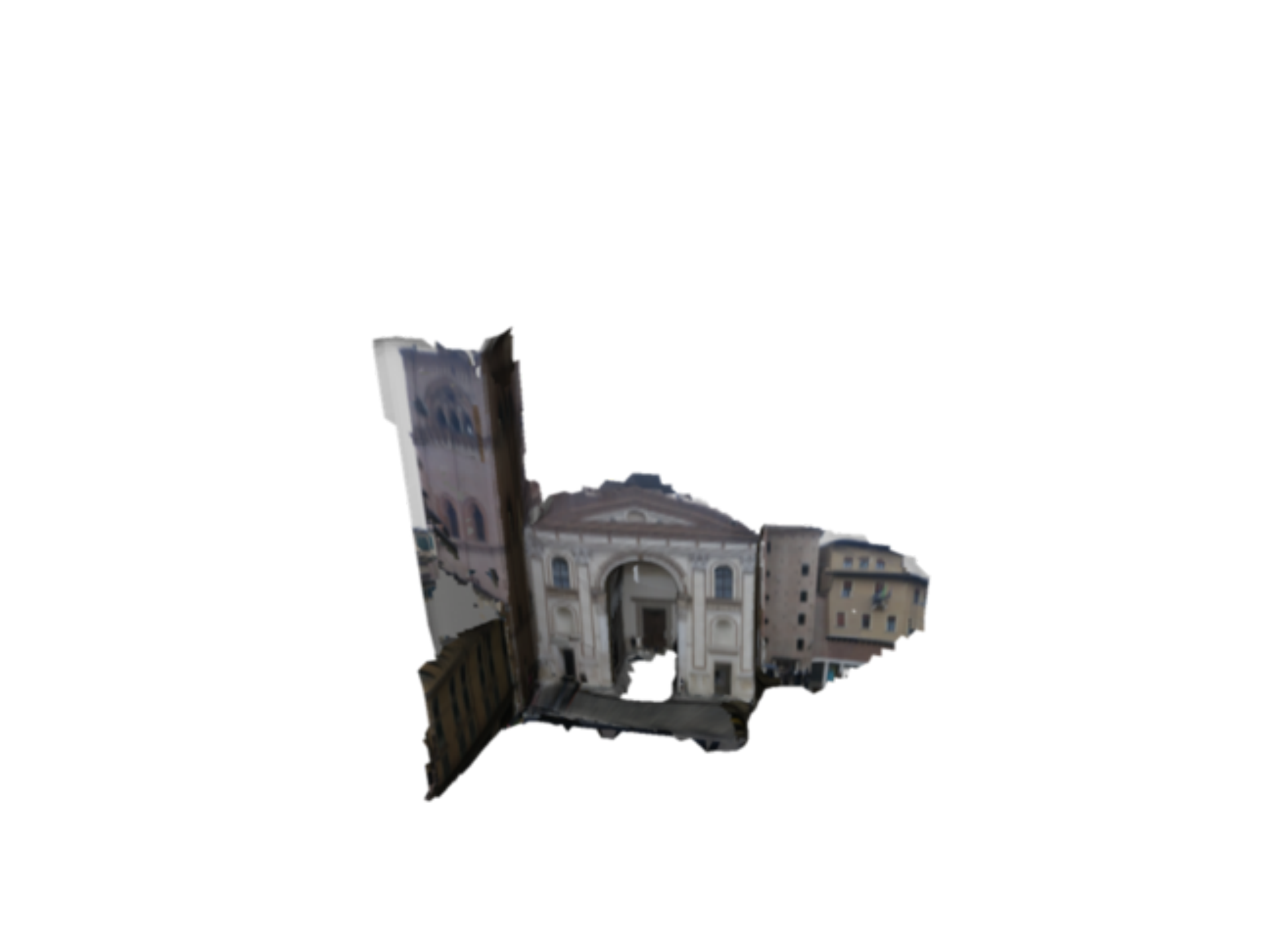} \\
		 \includegraphics[width=0.33\linewidth]{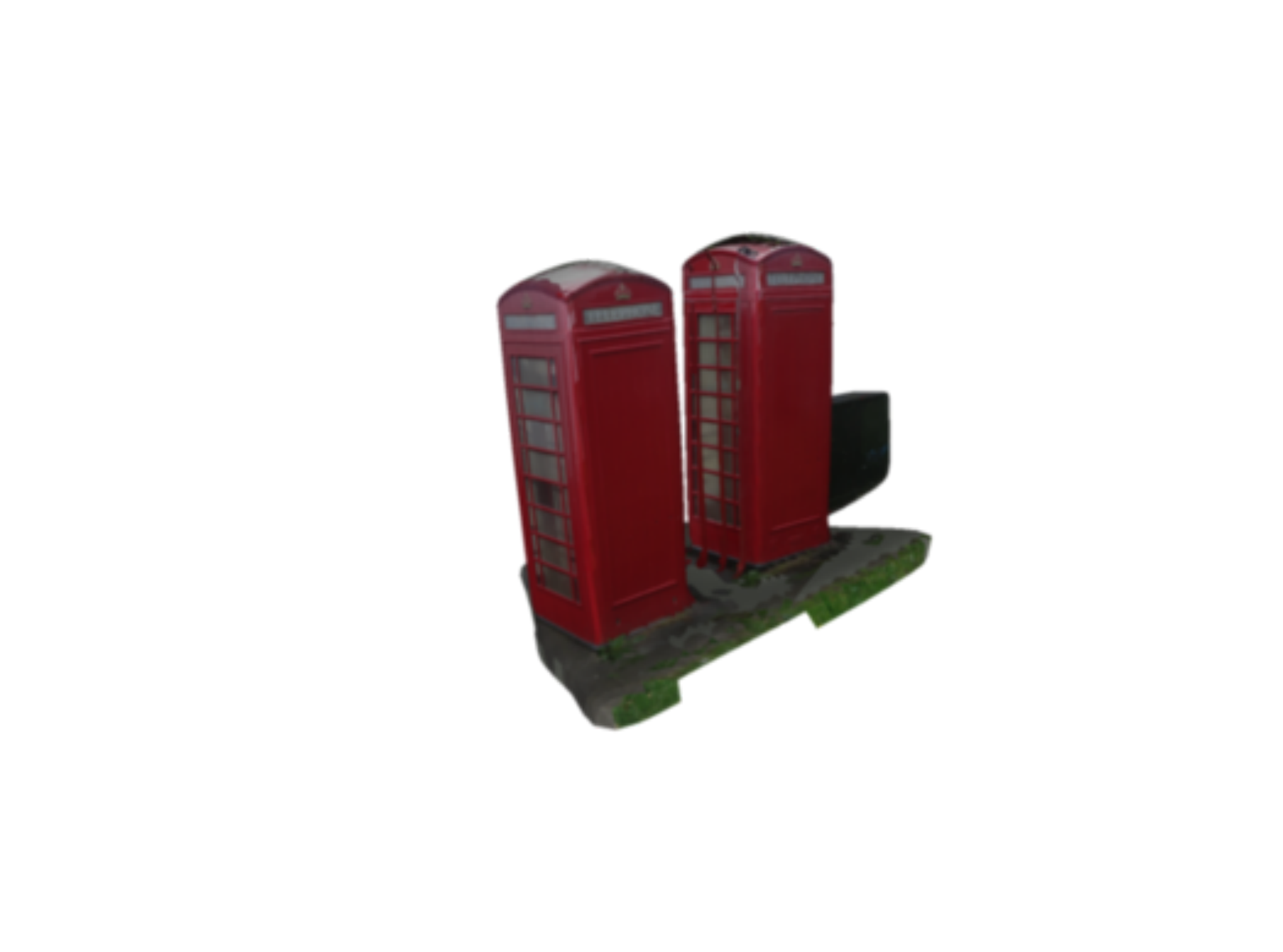} &
		 \includegraphics[width=0.33\linewidth]{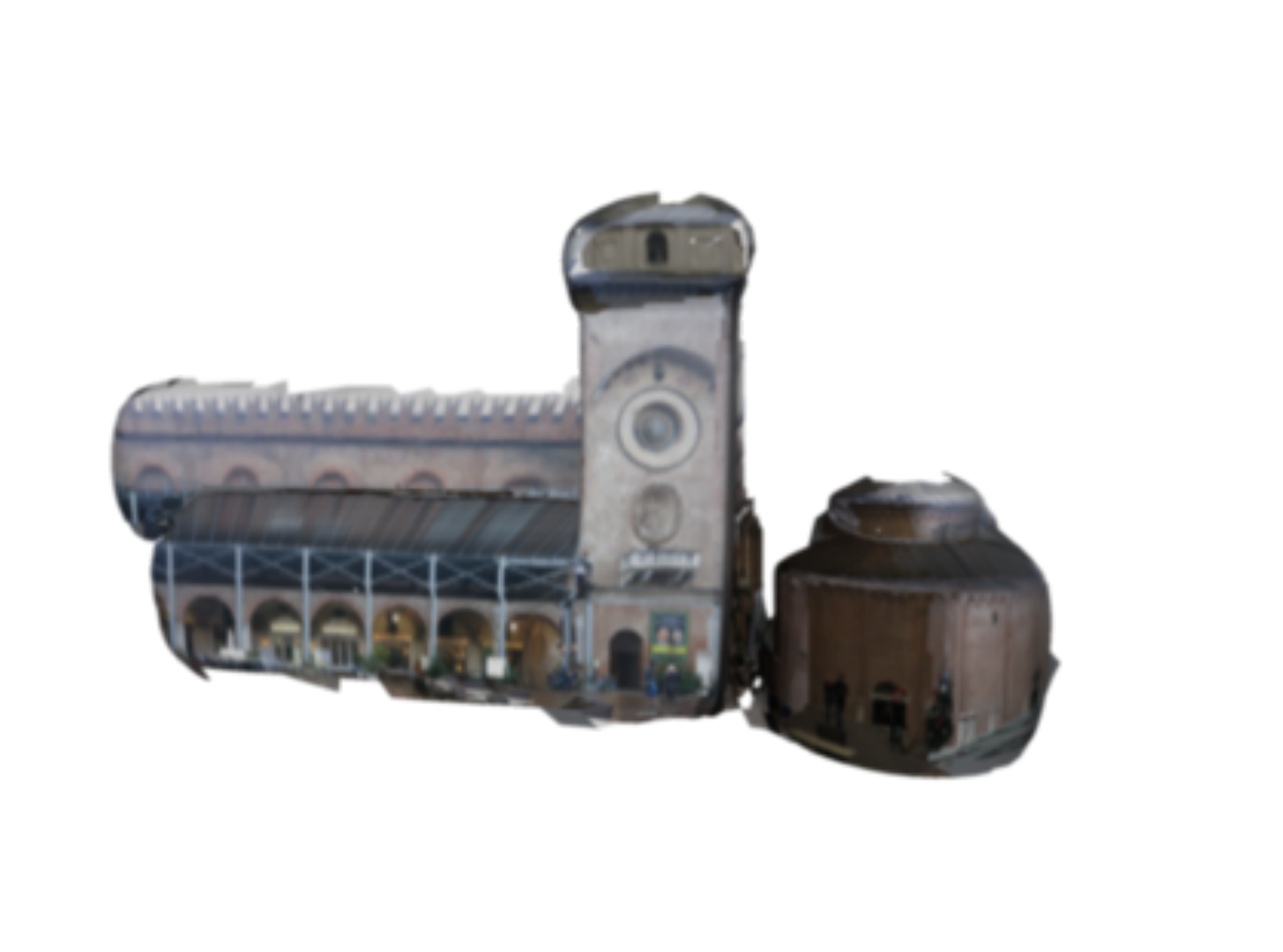} &
		 \includegraphics[width=0.33\linewidth]{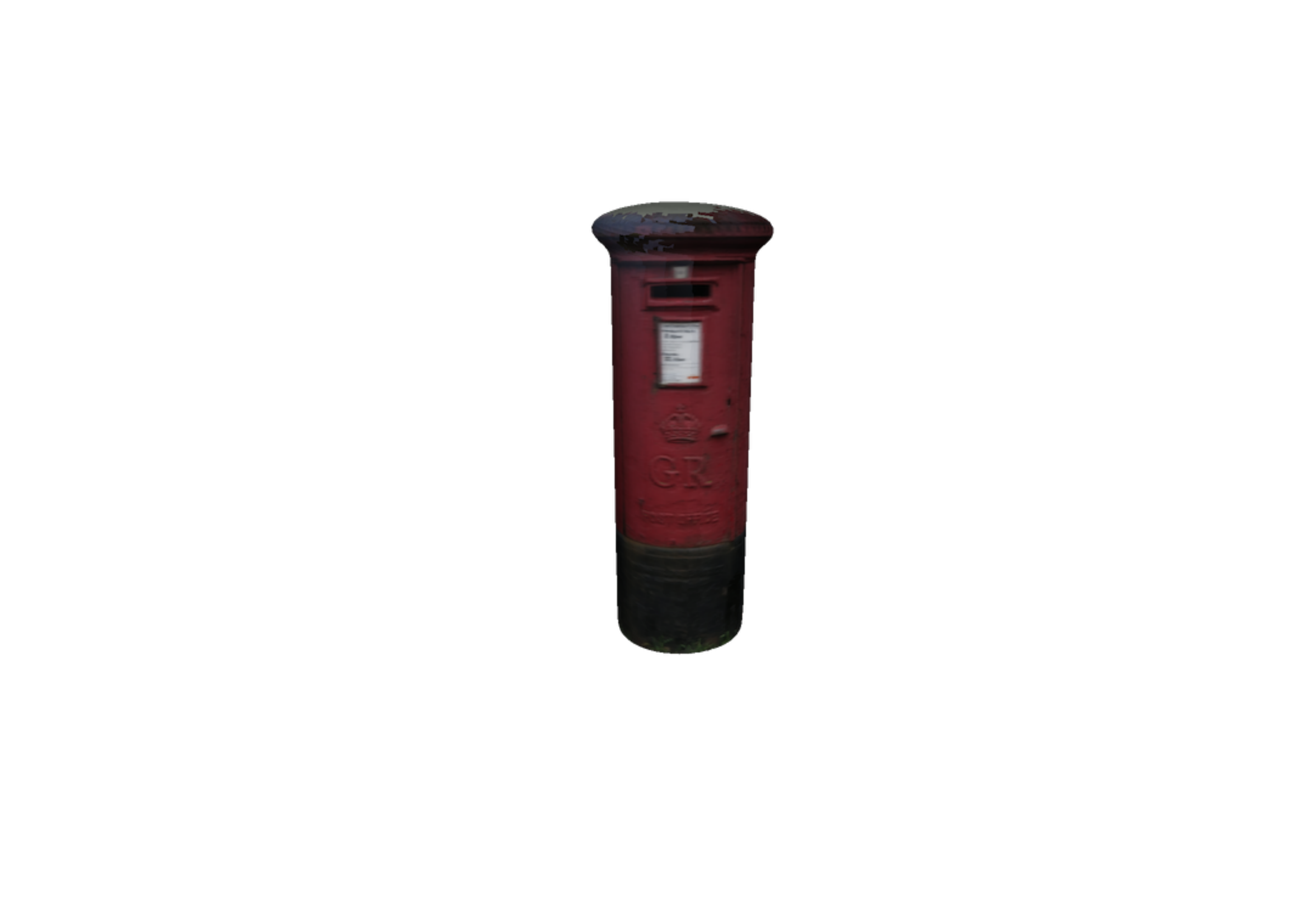}
	\end{tabular}
	\caption{The final textured results.}
	\label{fig:finalresultstextured}
\end{figure*}

The results can be visually compared in Figure \ref{fig:finalresults}. It can be seen that the Poisson reconstruction fails where there are large holes in the surface such as the floor of the post-office, or on the roof of the phone booths. The final textured meshes are shown in Figure \ref{fig:finalresultstextured}. 

The software is demonstrated in the supplementary material. A candidate user is recorded fitting primitives using only free-form input. The videos demonstrate the ease of use for both latent variable models and quadric surfaces, on a touch screen display. The use of Bayesian statistics makes the system robust to noise in the input trajectories, resulting in geometry which is a good fit to the data.

\section{Conclusion}
The system that has been presented in this paper enables a user to quickly fit parametric surfaces to point clouds. It is demonstrated that we outperform alternative methods on a dataset of medium to large scale objects such as buildings and street furniture, which are man-made and comprise surfaces which are piecewise primitive. Alternative interactive methods either do not make the correct assumptions about the data or do not correctly deal with occluded surfaces, making them laborious for desktop use by an artist. 

The system performs well on data which exhibits a large amount of noise, and has holes in the point cloud; a limiting artefact for modern autonomous reconstruction techniques such as Poisson reconstruction. 

While the system does perform well under the assumptions, it is not expected to work well if they are broken. Highly complicated natural objects such as flowing water, vegetation and so on, require an algorithm which does not constrain the smoothness of the surface to the same degree. To some extent, simplicity is a matter of scale, and so at a close range one might expect better results; but it is an extension that we do not consider in this work.

We conclude that the system provides a useful interface for reconstructing surfaces from urban scenes, and produces high quality and visually pleasing results.

\bibliographystyle{plainnat}
\bibliography{Bibliography} 

\end{document}